\documentclass[12pt]{article}
\usepackage{psfig}
\usepackage{float}
\usepackage{subeqn}
\usepackage{amssymb}
\ifx\pdfoutput\undefined
\usepackage{graphicx}
\else
\usepackage[pdftex]{graphicx}
\fi

\begin{document}

\begin{center}
{\Large \bf Carbon Nanotube Thin Film Field Emitting Diode:
Understanding the System Response Based on Multiphysics Modeling
\\}
\vspace{1cm}
{N. Sinha$^a$, D. Roy Mahapatra$^b$, 
J.T.W. Yeow$^a$\footnote{Corresponding author:
JTWY e-mail: jyeow@engmail.uwaterloo.ca;\\
Tel: 1 (519) 8884567 x 2152;
Fax: 1 (519) 7464791}, R.V.N. Melnik$^b$ and
D.A. Jaffray$^c$\\}
\vspace{0.5 cm}
{\it $^a$Department of Systems Design Engineering,
University of Waterloo, ON, N2L3G1, Canada\\
$^b$Mathematical Modeling and Computational Sciences,
Wilfrid Laurier University, Waterloo,
ON, N2L3C5, Canada\\
$^c$Department of Radiation Physics, Princess Margaret Hospital,
Toronto, ON, M5G2M9, Canada}
\end{center}

\centerline{\bf \large Abstract} In this paper, we model the evolution and self-assembly of
randomly oriented carbon nanotubes (CNTs), grown on a metallic substrate in the form of a
thin film for field emission under diode configuration. Despite high output, the current in
such a thin film device often decays drastically. The present paper is focused on
understanding this problem. A systematic, multiphysics based modelling approach is proposed.
First, a nucleation coupled model for degradation of the CNT thin film is derived, where the
CNTs are assumed to decay by fragmentation and formation of clusters. The random orientation
of the CNTs and the electromechanical interaction are then modeled to explain the
self-assembly. The degraded state of the CNTs and the electromechanical force are employed to
update the orientation of the CNTs. Field emission current at the device scale is finally
obtained by using the Fowler-Nordheim equation and integration over the computational cell
surfaces on the anode side. The simulated results are in close agreement with the
experimental results. Based on the developed model, numerical simulations aimed at
understanding the effects of various geometric parameters and their statistical features on
the device current history are reported.

\noindent {\bf Keywords:} Field emission, carbon nanotube, degradation, electrodynamics,
self-assembly.

\section{Introduction}
\label{sec:intro} The conventional mechanism used for electron emission is thermionic in
nature where electrons are emitted from hot cathodes (usually heated filaments). The
advantage of these hot cathodes is that they work even in environments that contain a large
number of gaseous molecules. However, thermionic cathodes in general have slow response time
and they consume high power. These cathodes have limited lifetime due to mechanical wear. In
addition, the thermionic electrons have random spatial distribution. As a result, fine
focusing of electron beam is very difficult. This adversely affects the performance of the
devices such as X-ray tubes. An alternative mechanism to extract electrons is field emission,
in which electrons near the Fermi level tunnel through the energy barrier and escape to the
vacuum under the influence of a sufficiently high external electric field. The field emission
cathodes have faster response time, consume less power and have longer life compared to
thermionic cathodes. However, field emission cathodes require ultra-high vacuum as they are
highly reactive to gaseous molecules during the field emission.

The key to the high performance of a field emission device is the
behavior of its cathode. In the past, the performance of cathode materials
such as spindt-type emitters and nanostructured diamonds for field
emission was studied by Spindt {\it et al.}$^1$ 
,
Gotoh {\it et al.}$^2$ 
, and Zhu$^3$ 
.
However, the spindt type emitters suffer from high manufacturing cost and
limited lifetime. Their failure is often caused by ion bombardment from the
residual gas species that blunt the emitter cones$^2$ 
.
On the other hand, nanostructured diamonds are unstable at high current
densities$^3$ 
. Carbon nanotube (CNT), which is an allotrope of
carbon, has potential to be used as cathode material in field emission devices.
Since their discovery by Iijima in 1991$^4$ 
, extensive
research on CNTs has been conducted.
Field emission from CNTs was first reported in 1995 by
Rinzler {\it et al.}$^5$ 
, de Heer {\it et al.}$^6$ 
, and Chernozatonskii {\it et al.}$^7$ 
.
Field emission from CNTs has been studied extensively since then.
Currently, with significant improvement in processing technique, CNTs
are among the best field emitters. Their applications in field
emission devices, such as field emission
displays, gas discharge tubes, nanolithography
systems, electron microscopes, lamps, and X-ray tube sources have
been successfully demonstrated$^{8-9}$ 
.
The need for highly controlled application of CNTs
in X-ray devices is one of the main reasons for the present study. The
remarkable field emission properties of CNTs are attributed to their geometry,
high thermal conductivity, and chemical stability.
Studies have reported that CNT sources have a high reduced brightness and
their energy spread values are comparable to conventional field emitters
and thermionic emitters$^{10}$ 
.

The physics of field emission from metallic surfaces is well understood. The
current density ($J$) due to field emission from a metallic surface is usually
obtained by using the Fowler-Nordheim (FN) equation$^{11}$ 
\begin{equation}
J=\frac {B E^2}{\Phi} \exp
\bigg(-\frac {C \Phi^{3/2}}{E}\bigg) \;,
\label{eq:damp1}
\end{equation}
where E is the electric field, $\Phi$ is the work function of the cathode material, and B and
C are constants. The device under consideration in this paper is a X-ray source where a thin
film of CNTs acts as the electron emitting surface (cathode). Under the influence of
sufficiently high voltage at ultra high vacuum, the electrons are extracted from the CNTs and
hit the heavy metal target (anode) to produce X-rays. However, in the case of a CNT thin film
 acting as cathode, the surface of the cathode is not smooth (like the metal emitters). In this case,
the cathode consists of hollow tubes grown on a substrate. Also, some amount of carbon
clusters may be present within the CNT-based film. An added complexity is that there is
realignment of individual CNTs due to electrodynamic interaction between the neighbouring
CNTs during field emission. At present, there is no adequate mathematical models to address
these issues. Therefore, the development of an appropriate mathematical modeling approach is
necessary to understand the behavior of CNT thin film field emitters.

\subsection{Role of various physical processes in the degradation
of CNT field emitter} 
\label{sec:mechanisms} 
Several studies have reported experimental
observations in favour of considerable degradation and failure of CNT cathodes. These studies
can be divided into two categories: (i) studies related to degradation of single nanotube
emitters$^{12-17}$ 
and (ii) studies related to degradation of CNT
thin films$^{18-23}$ 
. Dean {\it et al.}$^{13}$ 
found gradual decrease of field emission of single walled carbon nanotubes (SWNTs) due to
``evaporation'' when large field emitted current ($300nA$ to $2\mu A$) was extracted. It was
observed by Lim {\it et al.}$^{23}$ 
that CNTs are susceptible to damage by exposure
to gases such as oxygen and nitrogen during field emission. Wei {\it et al.}$^{14}$
observed that after field emission over $30$ minutes at field emission current between $50$
and $120nA$, the length of CNTs reduced by $10\%$. Wang  {\it et al.}$^{15}$
observed two
types of structural damage as the voltage was increased: a piece-by-piece and
segment-by-segment splitting of the nanotubes, and a layer-by-layer stripping process.
Occasional spikes in the current-voltage curves were observed by Chung {\it et
al.}$^{16}$ 
when the voltage was increased. Avouris {\it et al.}$^{17}$ 
found that the CNTs break down when subjected to high bias over a long period of time.
Usually, the breakdown process involves stepwise increases in the resistance. In the
experiments performed by the present authors, peeling of the film from the substrate was
observed at high bias. Some of the physics pertinent to these effects is known but the
overall phenomenon governing such a complex system is difficult to explain and quantify and
it requires further investigation.

There are several causes of CNT failures:
\begin{enumerate}
\item[(i)]
In case of multi-walled carbon nanotubes (MWNTs), the CNTs undergo
layer-by-layer stripping during field emission$^{15}$ 
.
The complete removal of the shells are most likely the reason for the
variation in the current voltage curves$^{16}$ 
;

\item[(ii)]
At high emitted currents, CNTs are
resistively heated. Thermal effect can sublime
a CNT causing cathode-initiated vacuum breakdown$^{24}$ 
.
Also, in case of thin films grown using chemical vapor deposition (CVD),
fewer catalytic metals such as nickel, cobalt, and iron are observed as
impurities in CNT thin films. These metal particles
melt and evaporate by high emission currents, and abruptly surge the emission
current. This results in vacuum breakdown followed by the failure of the CNT
film$^{23}$ 
;

\item[(iii)]
Gas exposure induces chemisorption and physisorption of gas molecules on the surface of CNTs.
In the low-voltage regime, the gas adsorbates remain on the surface of the emitters. On the
other hand, in the high-voltage regime, large emission currents resistively anneal the tips,
and the strong electric field on the locally heated tips promotes the desorption of gas
adsorbates from the tip surface. Adsorption of materials with high electronegativity hinders
the electron emission by intensifying the local potential barriers. Surface morphology can be
changed by an erosion of the cap of the CNT as the gases desorb reactively from the surface
of the CNTs$^{25}$ 
;

\item[(iv)]
CVD-grown CNTs tend to show more defects
in the wall as their radius increases. Possibly, there are rearrangements of
atomic structures (for example, vacancy migration) resulting in the
reduction of length of CNTs$^{14}$ 
. In addition, the presence of
defects may act as a centre for nucleation for
voltage-induced oxidation, resulting in electrical breakdown$^{16}$ 
;

\item[(v)]
As the CNTs grow perpendicular to the substrate, the contact area of CNTs with the substrate
is very small. This is a weak point in CNT films grown on planar substrates, and CNTs may
fail due to tension under the applied fields$^{20}$ 
. Small nanotube diameters and
lengths are an advantage from the stability point of view.
\end{enumerate}

Although the degradation and failure of single nanotube emitters can be either abrupt or
gradual, the degradation and failure of a thin film emitter with CNT cluster is mostly
gradual. The gradual degradation occurs either during initial current-voltage
measurement$^{21}$ 
 (at a fast time scale) or during measurements at constant
applied voltage over a long period of time$^{22}$ 
 (at a slow time scale).
Nevertheless, it can be concluded that the gradual degradation of thin films occurs due to
the failure of individual emitters.

Till date, several studies have reported experimental observations on CNT thin
films$^{26}$ 
. However, from mathematical, computational and design view points,
the models and characterization methods are available only for vertically aligned CNTs grown
on the patterned surface$^{27-28}$ 
. In a CNT film, the array of CNTs may
ideally be aligned vertically. However, in this case it is desired that the individual CNTs
be evenly separated in such a way that their spacing is greater than their height to minimize
the screening effect$^{29}$ 
. If the screening effect is minimized following the
above argument, then the emission properties as well as the lifetime of the cathodes are
adversely affected due to the significant reduction in density of CNTs. For the cathodes with
randomly oriented CNTs, the field emission current is produced by two types of sources: (i)
small fraction of CNTs that point toward the anode and (ii) oriented and curved CNTs
subjected to electromechanical forces causing reorientation. As often inferred (see e.g.,
ref.$^{29}$ 
), the advantage of the cathodes with randomly oriented CNTs is that
always a large number of CNTs take part in the field emission, which is unlikely in the case
of cathodes with uniformly aligned CNTs. Such a thin film of randomly oriented CNTs will be
considered in the present study. From the modeling point of view, its analysis becomes much
more challenging. Although some preliminary works have been reported (see e.g.,
refs.$^{30-31}$ 
), neither a detailed model nor a subsequent
characterization method are available that would allow to describe the array of CNTs that may
undergo complex dynamics during the process of charge transport. In the detailed model, the
effects of degradation and fragmentation of CNTs during field emission need to be considered.
However, in the majority of analytical and design studies, the usual practice is to employ
the classical Fowler-Nordheim equation$^{11}$ 
to determine the field emission from
the metallic surface, with correction factors to deal with the CNT tip geometry. Ideally, one
has to tune such an empirical approach to specific materials and methods used (e.g. CNT
geometry, method of preparation, CNT density, diode configuration, range of applied voltage,
 etc.). Also, in order to account for the oriented CNTs and interaction between themselves, it
is necessary to consider the space charge and the electromechanical forces. By taking into
account the evolution of the CNTs, a modeling approach is developed in this paper. In order
to determine phenomenologically the concentration of carbon clusters due to degradation of
CNTs, we introduce a homogeneous nucleation rate. This rate is coupled to a moment model for
the evolution. The moment model is incorporated in a spatially discrete sense, that is by
introducing volume elements or cells to physically represent the CNT thin film.
Electromechanical forces acting on the CNTs are estimated in time-incremental manner. The
oriented state of CNTs are updated using a mechanics based model. Finally, the current
density is calculated by using the details regarding the CNT orientation angle and the
effective electric field in the Fowler-Nordheim equation.

The remainder of this paper is organized as follows: in Sec. 2, a model is
proposed, which combines the nucleation coupled model for CNT
degradation with the electromechanical forcing model. Section 3 illustrates
the computational scheme. Numerical simulations and the comparison of
the simulated current-voltage characteristics with experimental results
are presented in Sec. 4.

\section{Model formulation}
\label{sec:model}

\begin{figure}
\centerline{\psfig{file=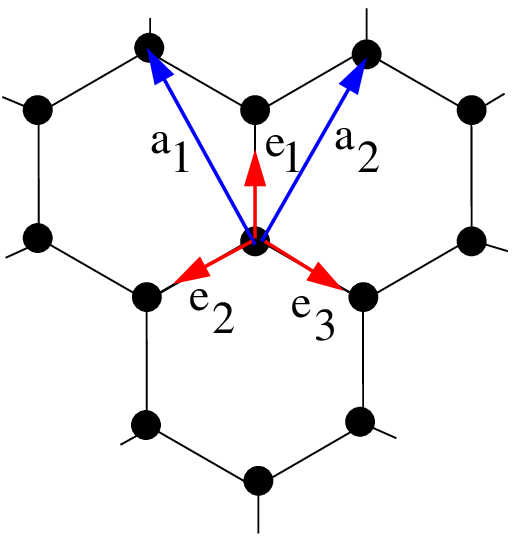,width=5cm}\hfill
            \psfig{file=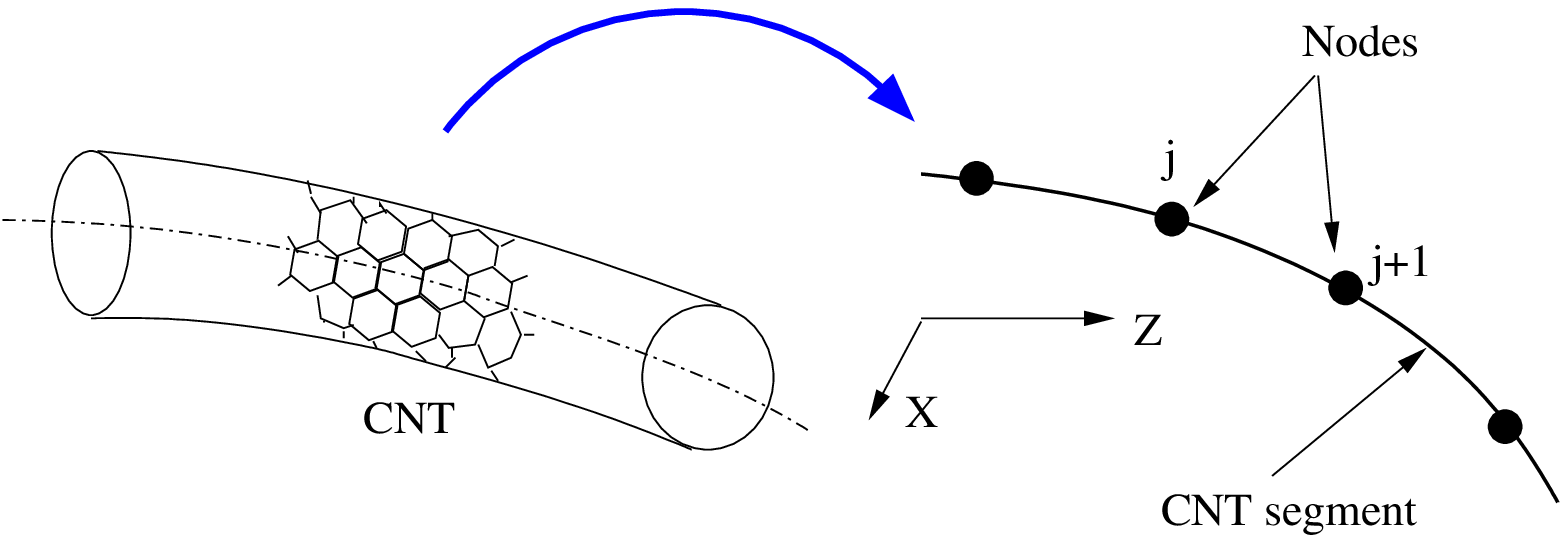,width=9cm}}
\centerline{\small(a)\hspace{7cm}(b)}
\caption{Schematic drawing showing (a) hexagonal arrangement of carbon atoms
in CNT and (b) idealization of CNT as a one-dimensional elastic member.}
\label{fig:film1}
\end{figure}

The CNT thin film is idealized in our mathematical model by using the following
simplifications.
\begin{enumerate}
\item[(i)]
CNTs are grown on a substrate to form a thin film. They are treated as
aggregate while deriving the nucleation coupled model for degradation
phenomenologically;

\item[(ii)]
The film is discretized into a number of representative volume element (cell), in which a
number of CNTs can be in oriented forms along with an estimated amount of carbon clusters.
This is schematically shown in Fig.~\ref{fig:film0}. The carbon clusters are assumed to be in
the form of carbon chains and networks (monomers and polymers);

\item[(iii)]
Each of the CNTs with hexagonal arrangement of carbon atoms (shown in
Fig.~\ref{fig:film1}(a)) are treated as effectively one-dimensional (1D) elastic members and
discretized by nodes and segments along its axis as shown in Fig.~\ref{fig:film1}(b).
Deformation of this 1D representation in the slow time scale defines the orientations of the
segments within the cell. A deformation in the fast time scale (due to electron flow) defines
the fluctuation of the sheet of carbon atoms in the CNTs and hence the resulting state of
atomic arrangements. The latter aspect is excluded from the present modeling and numerical
simulations, however they will be discussed within a quantum-hydrodynamic framework in a
forthcoming article.
\end{enumerate}

\subsection{Nucleation coupled model for degradation of CNTs}
\label{sec:evolution} Let $N_{T}$ be the total number of carbon atoms (in CNTs and in cluster
form) in a cell (see Fig.~\ref{fig:film0}). The volume of a cell is given by
$V_{\rm{cell}}=\Delta A d$, where $\Delta A$ is the cell surface interfacing the anode and
$d$ is distance between the inner surfaces of cathode substrate and the anode. Let $N$ be the
number of CNTs in the cell, and $N_{\rm{CNT}}$ be the total number of carbon atoms present in
the CNTs. We assume that during field emission some CNTs are decomposed and form clusters.
Such degradation and fragmentation of CNTs can be treated as the reverse process of CVD or a
similar growth process used for producing the CNTs on a substrate. Hence,
\begin{equation}
N_{T} =N N_{CNT}+ N_{\rm cluster} \;,
\label{eq:damp2}
\end{equation}
where $N_{\rm cluster}$ is the total number of carbon atoms in the
clusters in a cell at time $t$ and is given by
\begin{equation}
N_{\rm cluster} = V_{\rm cell} \int_{0}^{t} dn_{1}(t) \;,
\label{eq:damp3}
\end{equation}
where $n_{1}$ is the concentration of carbon cluster in the cell.
By combining Eqs.~(\ref{eq:damp2}) and (\ref{eq:damp3}),
one has
\begin{equation}
N =\frac {1} {N_{CNT}} \bigg [N_{T}-V_{\rm cell}
\int_{0}^{t} dn_{1}(t)\bigg] \;.
\label{eq:damp4}
\end{equation}
The number of carbon atom in a CNT is proportional to its length. Let the length of a CNT be
a function of time, denoted as $L(t)$. Therefore, one can write
\begin{equation}
N_{CNT}= N_{\rm ring} L(t) \;,
\label{eq:damp5}
\end{equation}
where $N_{\rm ring}$ is the number of carbon atoms per unit length of a CNT
and can be determined from the geometry of the hexagonal arrangement of carbon
atoms in the CNT.
By combining Eqs.~(\ref{eq:damp4}) and (\ref{eq:damp5}), one can write
\begin{equation}
N =\frac {1} {N_{\rm ring} L(t)} \bigg [N_{T}-V_{\rm cell}
\int_{0}^{t} dn_{1}(t)\bigg] \;.
\label{eq:damp6}
\end{equation}

In order to determine $n_{1}(t)$ phenomenologically, we need to know the nature of evolution
of the aggregate in the cell. From the physical point of view, one may expect the rate of
formation of the carbon clusters from CNTs to be a function of thermodynamic quantities, such
as temperature ($T$), the relative distances ($r_{ij}$) between the carbon atoms in the CNTs,
the relative distances between the clusters and a set of parameters ($p^*$) describing the
critical cluster geometry. The relative distance $r_{ij}$ between carbon atoms in CNTs is a
function of the electromechanical forces. Modeling of this effect is discussed in
Sec.~\ref{sec:orientation}. On the other hand, the relative distances between the clusters
influence in homogenizing the thermodynamic energy, that is, the decreasing distances between
the clusters (hence increasing densities of clusters) slow down the rate of degradation and
fragmentation of CNTs and lead to a saturation in the concentration of clusters in a cell.
Thus, one can write
\begin{equation}
\frac{d n_1}{dt} = f(T, r_{ij}, p^*) \;.
\label{eq:nucl0}
\end{equation}
To proceed further, we introduce a nucleation coupled model$^{32-33}$
,
which was originally proposed to simulate aerosol formation. Here we modify this model
according to the present problem which is opposite to the process of growth of CNTs from the
gaseous phase. With this model the relative distance function is replaced by a collision
frequency function ($\beta_{ij}$) describing the frequency of collision between the $i$-mers
and $j$-mers, with
\begin{equation}
\beta_{ij}=\bigg (\frac {3v_{1}}{4\pi}\bigg)^{1/6}
\sqrt {\frac {6kT} {\rho_{p}}\bigg
(\frac{1}{i}+\frac{1}{j}\bigg)}\bigg(i^{1/3}+j^{1/3}\bigg)^{2} \;,
\label{eq:damp14}
\end{equation}
and the set of parameters describing the critical cluster geometry by
\begin{equation}
p^* = \{ v_j\; s_j\; g^*\; d_p^* \} \;,
\end{equation}
where $v_j$ is the $j$-mer volume, $s_j$ is the surface area of $j$-mer,
$g^*$ is the normalized critical cluster size, $d_p^*$ is the critical
cluster diameter, $k$ is the Boltzmann constant, $T$ is the temperature and
$\rho_{p}$ is the particle mass density. 
The detailed form of Eq.~(\ref{eq:nucl0}) is given
by four nonlinear ordinary differential equations:
\begin{equation}
\frac {dN_{\rm kin}}{dt}= J_{\rm kin} \;,
\label{eq:damp7}
\end{equation}
\begin{equation}
\frac {dS}{dt}=-\frac {J_{\rm kin} S g^{*}} {n_{1}} - (S-1)
\frac {B_{1} A_{n}} {2v_{1}} \;,
\label{eq:damp8}
\end{equation}
\begin{equation}
\frac {dM_{1}}{dt}=J_{\rm kin}d_{p}^{*} + (S-1) B_{1}N_{\rm kin} \;,
\label{eq:damp9}
\end{equation}
\begin{equation}
\frac {dA_{n}}{dt}=\frac {J_{\rm kin}S{g^{*}}^ {2/3} s_{1}} {n_{1}}
+ \frac {2 \pi B_{1}S(S-1) M_{1}} {n_{1}} \;,
\label{eq:damp10}
\end{equation}
where $N_{\rm kin}$ is the kinetic normalization constant,
$J_{\rm kin}$ is the kinetic nucleation rate, $S$ is the saturation ratio,
$A_{n}$ is the total surface area of the carbon cluster and
$M_{1}$ is the moment of cluster size distribution. The quantities involved
are expressed as
\begin{equation}
S=\frac{n_{1}}{n_{s}} \;, \quad
M_{1}=\int_{d_{p}^{*}}^{d_{p}^{\max}} \Big(n(d_{p}, t) d_{p}\Big) d(d_{p}) \;,
\label{eq:damp11}
\end{equation}
\begin{equation}
N_{\rm kin}=\frac {n_{1}}{S} \exp(\Theta) \;, \quad
J_{\rm kin}=\frac {\beta_{ij}n_{1}^{2}}{12S}
\sqrt \frac {\Theta} {2\pi} \exp {\bigg (\Theta - \frac{4\Theta ^ {3}}{27
(\ln{S})^{2}}\bigg)} \;,
\label{eq:damp12}
\end{equation}
\begin{equation}
g^{*}=\bigg (\frac {2}{3} \frac {\Theta}{\ln S}\bigg)^{3} \;,
\quad d_{p}^{*}=\frac {4\sigma v_{1}}{kT\ln S} \;,
\quad B_{1}=2n_{s}v_{1}\sqrt \frac {kT}{2 \pi m_{1}} \;,
\label{eq:damp13}
\end{equation}
where $n_{s}$ is the equilibrium saturation concentration of carbon cluster, $d_{p}^{\max}$
is the maximum diameter of the clusters, $n(d_{p},t)$ is the cluster size distribution
function, $d_{p}$ is the cluster diameter, $m_j$ is the mass of $j$-mer, $\Theta$ is the
dimensionless surface tension given by
\begin{equation}
\Theta=\frac {\sigma s_{1}}{kT} \;,
\label{eq:damp15}
\end{equation}
$\sigma$ is the surface tension. In this paper, we have considered $i=1$ and $j=1$
for numerical simulations, that is, only monomer type clusters are considered. In
Eqs.~(\ref{eq:damp7})-(\ref{eq:damp10}), the variables are $n_1(t)$, $S(t)$, $M_1(t)$ and
$A_n(t)$, and all other quantities are assumed constant over time. In the expression for
moment $M_1(t)$ in Eq.~(\ref{eq:damp11}), the cluster size distribution in the cell is
assumed to be Gaussian, however, random distribution can be incorporated. We solve
Eqs.~(\ref{eq:damp7})-(\ref{eq:damp10}) using a finite difference scheme as discussed in
Sec.~\ref{sec:computation}. Finally, the number of CNTs in the cell at a given time is
obtained with the help of Eq.~(\ref{eq:damp6}), where the reduced length $L(t)$ is determined
using geometric properties of the individual CNTs as formulated next.

\subsection{Effect of CNT geometry and orientation}
\label{sec:orientation}
It has been discussed in Sec.~\ref{sec:mechanisms} that the
geometry and orientation of the tip of the CNTs are important factors in the
overall field emission performance of the film and must be considered in the
model.

\begin{figure}
\centerline{\psfig{file=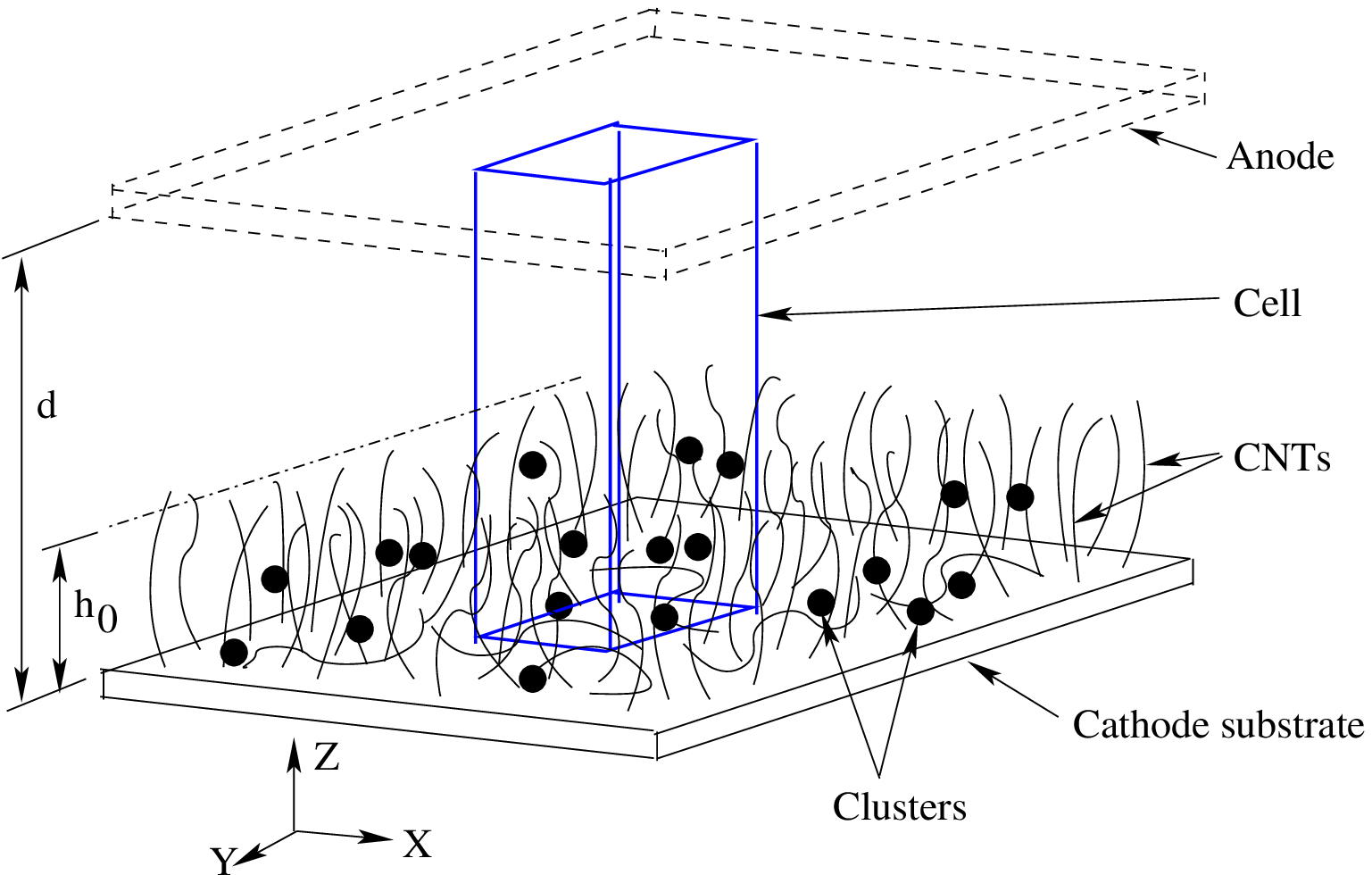,width=8cm}}
\caption{Schematic drawing of the CNT thin film for model idealization.}
\label{fig:film0}
\end{figure}

As an initial condition, let $L(0)=h$ at $t=0$, and let
$h_0$ be the average height of the CNT region as shown
in Fig.~\ref{fig:film0}. This average height $h_0$ is approximately
equal to the height of the CNTs that are aligned vertically.
If $\Delta h$ is the decrease in the length of a CNT (aligned vertically
or oriented as a segment) over a time interval $\Delta t$ due to
degradation and fragmentation, and if $d_{t}$ is the diameter of the CNT,
then the surface area of the CNT decreased is $\pi d_{t} \Delta h$.
By using the geometry of the CNT, the decreased
surface area can be expressed as
\begin{equation}
\pi d_{t} \Delta h
=V_{\rm cell} n_{1}(t){\bigg [s(s-a_{1})(s-a_{2})(s-a_{3})\bigg]}^{1/2} \;,
\label{eq:damp16}
\end{equation}
where $V_{\rm{cell}}$ is the volume of the cell as introduced in Sec.~\ref{sec:evolution},
$a_{1}$, $a_{2}$, $a_{3}$ are the lattice constants, and $s=\frac {1}{2}(a_{1}+ a_{2}+a_{3})$
(see Fig.~\ref{fig:film1}(a)). The chiral vector for the CNT is expressed as
\begin{equation}
\overrightarrow C_{h}= n \vec a_{1} + m \vec a_{2} \;,
\label{eq:damp161}
\end{equation}
where $n$ and $m$ are integers $(n \geq |m| \geq 0)$ and the pair
$(n,m)$ defines the chirality of the CNT. The following properties hold:
$\vec a_{1}.\vec a_{1}=a_{1}^{2}$, $\vec a_{2}.\vec a_{2}=a_{2}^{2}$,
and $2\vec a_{1}.\vec a_{2}=a_{1}^{2}+a_{2}^{2}-a_{3}^{2}$. With the help
of these properties the circumference and the diameter of the CNT can be
expressed as, respectively$^{34}$
,
\begin{equation}
|\overrightarrow C_{h}|=\sqrt {n^{2} a_{1}^{2}+ m^{2} a_{2}^{2} +
nm(a_{1}^{2}+ a_{2}^{2}-a_{3}^{2})} \;, \quad
d_{t}= \frac {|\overrightarrow C_{h}|}{\pi} \;,
\label{eq:damp163}
\end{equation}
Let us now introduce the rate of degradation of the CNT or simply the burning rate as
$\displaystyle v_{\rm burn}= \lim_{\Delta t \rightarrow 0} \Delta h/\Delta t$. By dividing
both side of Eq.~(\ref{eq:damp16}) by $\Delta t$ and by applying limit, one has
\begin{equation}
\pi d_{t}v_{\rm burn}
=V_{\rm cell} \frac {dn_{1}(t)}{dt}
{\bigg [s(s-a_{1})(s-a_{2})(s-a_{3})\bigg]}^{1/2} \;,
\label{eq:damp17}
\end{equation}
By combining Eqs.~(\ref{eq:damp163}) and (\ref{eq:damp17}), the burning rate
is finally obtained as
\begin{equation}
v_{\rm burn}=V_{\rm cell} \frac {dn_{1}(t)}{dt}
\bigg[\frac {s(s-a_{1})(s-a_{2})(s-a_{3})}
{n^2a_{1}^2+m^2a_{2}^2+nm(a_{1}^2+a_{2}^2-a_{3}^2)} \bigg]^{1/2} \;.
\label{eq:damp18}
\end{equation}

\begin{figure}
\centerline{\psfig{file=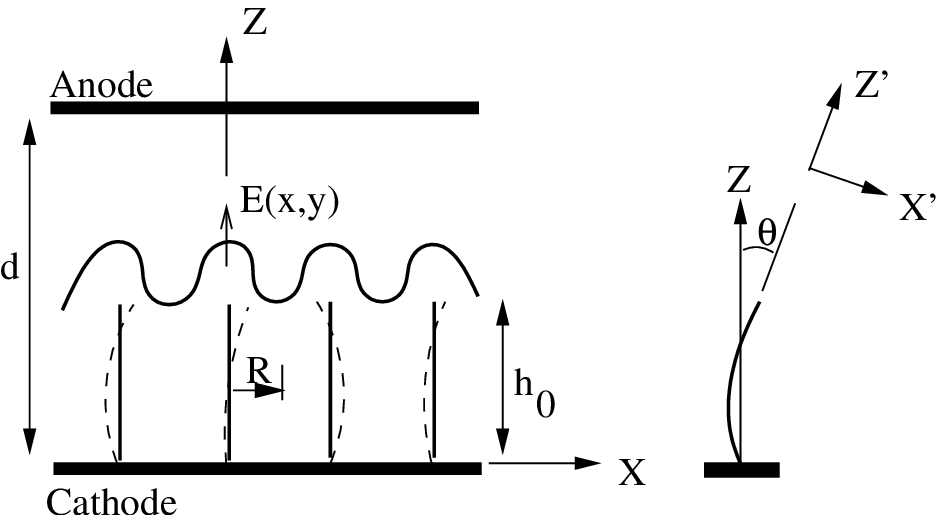,width=7cm}}
\caption{CNT array configuration.}
\label{fig_electric}
\end{figure}

In Fig.~\ref{fig_electric} we show a schematic drawing of
the CNTs almost vertically aligned, that is along the direction of
the electric field $E(x,y)$. This electric field $E(x,y)$ is assumed to be
due to the applied bias voltage. However, there will be an
additional but small amount of electric field due to several localized
phenomena (e.g., electron flow in curved CNTs, field emission from
the CNT tip etc.). Effectively, we assume that
the distribution of the field parallel to $z$-axis is of periodic nature
(as shown in Fig.~\ref{fig_electric}) when the CNT tips are vertically
oriented. Only a cross-sectional view in the $xz$ plane is shown in
Fig.~\ref{fig_electric} because only an array of CNTs
across $x$-direction will be considered in the model for simplicity. Thus,
in this paper, we shall restrict our attention to a two-dimensional problem,
and out-of-plane motion of the CNTs will not be incorporated in the model.

To determine the effective electric field at the tip of a CNT oriented at an angle $\theta$
as shown in Fig.~\ref{fig_electric}, we need to know the tip coordinate with respect to the
cell coordinate system. If it is assumed that a CNT tip was almost vertically aligned at
$t=0$ (as it is the desired configuration for the ideal field emission cathode), then its
present height is $L(t)=h_0-v_{\rm{burn}}t$ and the present distance between the tip and the
anode is $d_g=d-L(t)=d-h_0+v_{\rm{burn}}t$. We assume that the tip electric field has a
$z$-dependence of the form $E_0 L(t)/d_g$, where $E_0=V/d$ and $V$ is the applied bias
voltage. Also, let ($x,y$) be the deflection of the tip with respect to its original location
and the spacing between the two neighboring CNTs at the cathode substrate is $2R$. Then the
electric field at the deflected tip can be approximated as
\begin{equation}
E_{z'}=\sqrt {1-\frac {x^2+y^2}{R^2}}
\frac{(h_0-v_{\rm{burn}}t)}{(d-h_0+v_{\rm{burn}}t)} E_0 \;, \quad
\theta(t) \le \theta_c \;,
\label{eq:damp19}
\end{equation}
where $\theta_c$ is a critical angle to be set during numerical calculations along with the
condition: $E_{z'}=0$ when $\theta(t) > \theta_c$. This is consistent with the fact that
those CNTs which are low lying on the substrate do not contribute to the field emission. The
electric field at the individual CNT tip derived here is defined in the local coordinate
system ($X',Z'$) as shown in Fig.~\ref{fig_electric}. The components of the electric field in
the cell coordinate system ($X,Y,Z$) is given by the following transformation:
\begin{equation}
\left[ \begin{array}{ccc}
E_z\\ E_x\\ E_y\\
\end{array} \right]
=\left[ \begin{array}{ccc}
n_{z} & l_{z} & m_{z} \\
\sqrt {1-n_{z}^{2}} & \frac {-l_{z}n_{z}}{\sqrt {1-n_{z}^{2}}} & \frac {m_{z}n_{z}}{\sqrt {1-n_{z}^{2}}} \\
0 & \frac {-l_{z}n_{z}}{\sqrt {1-n_{z}^{2}}} & \frac {-l_{z}}{\sqrt {1-n_{z}^{2}}} \\
\end{array} \right]
\left[ \begin{array}{ccc}
E_{z'} \\ 0 \\ 0 \\
\end{array} \right] \;,
\label{eq:damp23}
\end{equation}
where $n_{z}$, $l_{z}$, $m_{z}$ are the direction cosines. According to the
cell coordinate system in Figs.~\ref{fig:film0} and \ref{fig_electric},
$n_{z}= \cos \theta(t)$, $l_{z}= \sin \theta(t)$, and $m_{z}= 0$.
Therefore, Eq.~(\ref{eq:damp23}) can be rewritten as
\begin{equation}
\left[ \begin{array}{ccc}
E_{z}\\ E_{x}\\ E_{y}\\
\end{array} \right]
=\left[ \begin{array}{ccc}
\cos \theta (t) & \sin \theta (t) & 0 \\
\sqrt {1-\cos^{2} \theta (t)} & -\cos \theta (t) & 0 \\
0 & -\cos \theta (t) & -1 \\
\end{array} \right]
\left[ \begin{array}{ccc}
E_{z'} \\ 0 \\ 0 \\
\end{array} \right] \;.
\label{eq:damp24}
\end{equation}
By simplifying Eq.~(\ref{eq:damp24}), we get
\begin{equation}
E_z=E_{z'} \cos \theta(t) \;, \quad
E_x=E_{z'} \sin \theta(t) \;.
\label{eq:damp25}
\end{equation}
Note that the identical steps of this transformation also apply to a generally oriented
($\theta\ne 0$) segment of CNT as idealized in Fig.~\ref{fig:film1}(b). 
The electric field components $E_z$
and $E_x$ are later used for calculation of the electromechanical force acting on the CNTs.
Since in this study we aim at estimating the current density at the anode due to the field
emission from the CNT tips, we also use $E_{z}$ from Eq.~(\ref{eq:damp25}) to compute the
output current based on the Fowler-Nordheim equation~(\ref{eq:damp1}).

\subsection{Electromechanical forces}
\label{sec:electromechanical} For each CNT, the angle of orientation $\theta(t)$ is dependent
on the electromechanical forces. Such dependence is geometrically nonlinear and it is not
practical to solve the problem exactly, especially in the present situation where a large
number of CNTs are to be dealt with. However, it is possible to solve the problem in
time-dependent manner with an incremental update scheme. In this section we derive the
components of the electromechanical forces acting on a generally oriented CNT segment. The
numerical solution scheme based on an incremental update scheme will be discussed in
Sec.~\ref{sec:computation}.

From the studies reported in published literature and based on the discussions made in
Sec.~\ref{sec:mechanisms}, it is reasonable to expect that the major contribution is due to
(i) the Lorentz force under electron gas flow in CNTs (a hydrodynamic formalism), (ii) the
electrostatic force (background charge in the cell), (iii) the van der Waals force against
bending and shearing of MWNT and (iv) the ponderomotive force acting on the CNTs.

\subsubsection{Lorentz force}
It is known that the electrical conduction and related properties of CNTs depend on the
mechanical deformation and the geometry of the CNT. In this paper we model the field emission
behaviour of the CNT thin film by considering the time-dependent electromechanical effects,
whereas the electronic properties and related effects are incorporated through the
Fowler-Nordheim equation empirically. Electronic band-structure calculations are
computationally prohibitive at this stage and at the same spatio-temporal scales considered
for this study. However, a quantum-hydrodynamic formalism seems practical and such details
will be dealt in a forthcoming article. Within the quantum-hydrodynamic formalism, one
generally assumes the flow of electron gas along the cylindrical sheet of CNTs. The
associated electron density distribution is related to the energy states along the length of
the CNTs including the tip region. What is important for the present modeling is that the
CNTs experience Lorentz force under the influence of the bias electric field as the electrons
flow from the cathode substrate to the tip of a CNT. The Lorentz force is expressed as
\begin{equation}
\vec f_{l}= e (\hat{n}_{0}+\hat{n}_{1}) \vec E \approx e \hat{n}_{0} \vec E \;,
\label{eq:damp26}
\end{equation}
where $e$ is the electronic charge, $\hat{n}_{0}$ is the surface electron
density corresponding to the Fermi level energy, $\hat{n}_{1}$ is the
electron density due to the deformation in the slow time scale, and
phonon and electromagnetic wave coupling at the fast time scale, and $\vec E$ is the
electric field. The surface electron density corresponding to the Fermi
level energy is expressed as$^{35}$ 
\begin{equation}
\hat{n}_{0}= \frac {kT}{\pi b^{2} \Delta} \;,
\label{eq:damp27}
\end{equation}
where $b$ is the interatomic distance and $\Delta$ is the overlap integral ($\approx 2 eV$
for carbon). The quantity $b$ can be related to the mechanical deformation of the 1D segments
(See Fig.~\ref{fig:film1}) and formulations reported by Xiao {\it et al.}$^{36}$
can
be employed. For simplicity, the electron density fluctuation $\hat{n}_{1}$ is neglected 
in this paper. Now, with the electric field components derived in
Eq.~\ref{eq:damp25}, the components of the Lorentz force acting along $z$ and $x$ directions
can now be written as, respectively,
\begin{equation}
f_{lz}= \pi d_{t} e \hat{n}_{0} E_z \;, \quad
f_{lx}= \pi d_{t} e \hat{n}_{0} E_x \approx 0  \;.
\label{eq:damp29}
\end{equation}

\subsubsection{Electrostatic force}
\begin{figure}
\centerline{\psfig{file=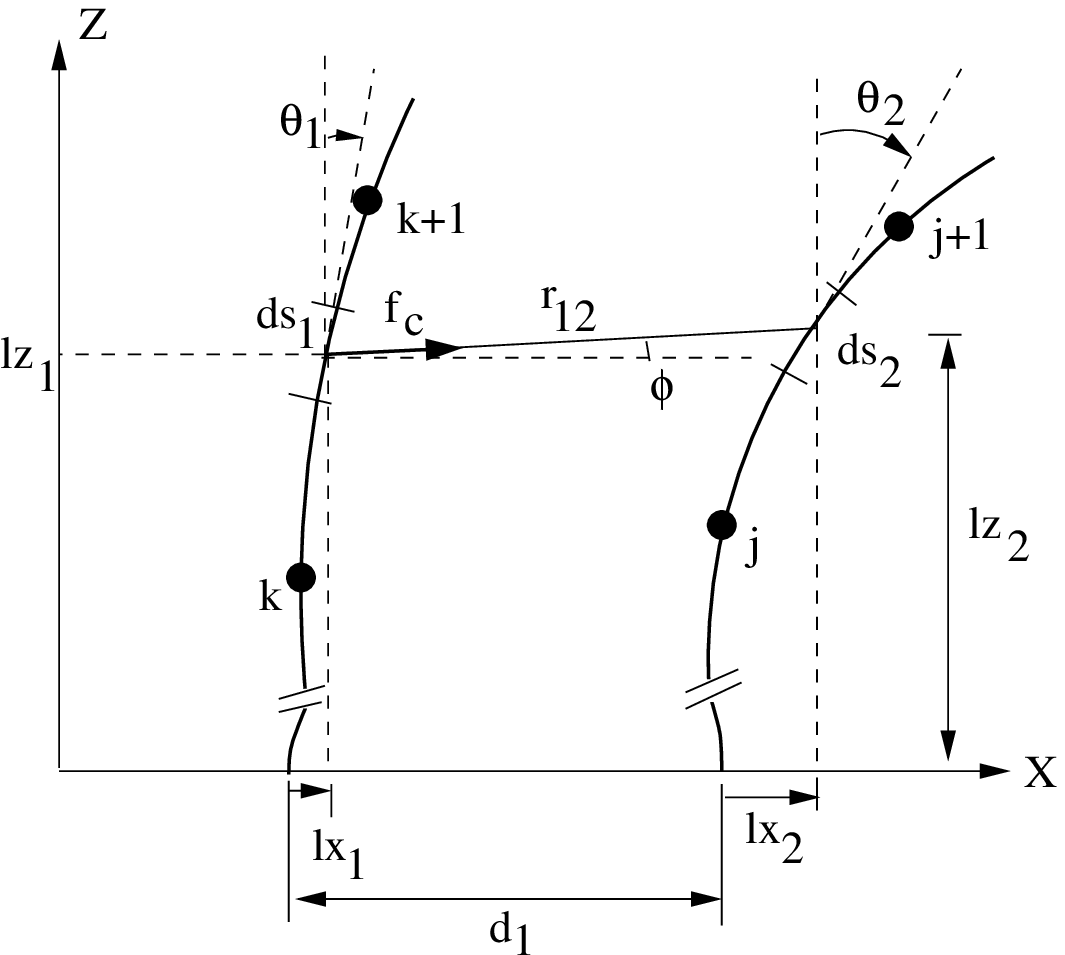,width=7cm}}
\caption{Schematic description of neighboring CNT pair interaction for
calculation of electrostatic force.}
\label{fig_projection}
\end{figure}

In order to calculate the electrostatic force, the interaction among two neighboring CNTs is
considered. For such calculation, let us consider a segment $ds_{1}$ on a CNT (denoted 1) and
another segment $ds_{2}$ on its neighboring CNT (denoted 2). These are parts of the
representative 1D member idealized as shown in Fig.~\ref{fig:film1}(b). The charges
associated with these two segments can be expressed as
\begin{equation}
q_{1}= e \hat{n}_{0} \pi d_{t}^{(1)} ds_{1} \;, \quad
q_{2}= e \hat{n}_{0} \pi d_{t}^{(2)} ds_{2} \;,
\label{eq:damp30}
\end{equation}
where $d_{t}^{(1)}$ and $d_{t}^{(2)}$ are diameters of two neighbouring
CNTs (1) and (2). The electrostatic force on the segment $ds_{1}$ by the
segment $ds_{2}$ is
\[
\frac {1}{4 \pi \epsilon \epsilon_{0}} \frac {q_{1}q_{2}}{r_{12}^2} \;,
\]
where $\epsilon$ is the effective permittivity of the aggregate of CNTs
and carbon clusters, $\epsilon_{0}$ is the permittivity of free space,
and $r_{12}$ is the effective distance between the
centroids of $ds_{1}$ and $ds_{2}$. The electrostatic force on the
segment $ds_{1}$ due to charge in the entire segment ($s_{2}$)
of the neighboring CNT (see Fig.~\ref{fig_projection}) can be written as
\[
\frac {1}{4 \pi \epsilon \epsilon_{0}} \int_{0}^{s_2}
\frac{1}{r_{12}^2} \left( e \hat{n}_{0} \pi d_{t}^{(1)} ds_{1}
e \hat{n}_{0} \pi d_{t}^{(2)} \right) ds_{2} \;.
\]
The electrostatic force per unit length on $s_1$ due to $s_{2}$ is then
\begin{equation}
f_{c}= \frac {1}{4 \pi \epsilon \epsilon_{0}} \int_{0}^{s_2}
\frac {(\pi e \hat{n}_{0})^2  d_{t}^{(1)} d_{t}^{(2)}}{r_{12}^2}
\ ds_{2} \;.
\label{eq:damp31}
\end{equation}
The differential of the force $df_c$ acts along the line joining the centroids
of the segments $ds_1$ and $ds_2$ as shown in Fig.~\ref{fig_projection}.
Therefore, the components of the total electrostatic force per
unit length of CNT (1) in $X$ and $Z$ directions
can be written as, respectively,
\[
f_{c_{x}}= \int df_{c} \cos \phi =
\frac {1}{4 \pi \epsilon \epsilon_{0}}
\int_{0}^{s_{2}} \frac {(\pi e \hat{n}_{0})^2  d_{t}^{(1)} d_{t}^{(2)}}
{r_{12}^2} \cos\phi \ ds_{2}
\]
\begin{equation}
\quad \equiv
\frac {1}{4 \pi \epsilon \epsilon_{0}} \sum_{j=1}^{h_0/\Delta s_2}
\frac {(\pi e \hat{n}_{0})^2  d_{t}^{(1)} d_{t}^{(2)}}{r_{12}^2} \
\cos \phi \ \Delta s_{2} \;,
\label{eq:damp32}
\end{equation}
\[
f_{c_{z}}= \int df_{c} \sin \phi =
\frac {1}{4 \pi \epsilon \epsilon_{0}} \int_{0}^{s_{2}}
\frac {(\pi e \hat{n}_{0})^2  d_{t}^{(1)} d_{t}^{(2)}}{r_{12}^2}
\sin \phi \ ds_2
\]
\begin{equation}
\quad \equiv \frac {1}{4 \pi \epsilon \epsilon_{0}}
\sum_{j=1}^{h_0/\Delta s_2}
\frac {(\pi e \hat{n}_{0})^2  d_{t}^{(1)} d_{t}^{(2)}}{r_{12}^2} \
\sin \phi \ \Delta s_{2} \;,
\label{eq:damp33}
\end{equation}
where $\phi$ is the angle the force vector $df_{c}$ makes with the $X$-axis.
For numerical computation of the above integrals, we compute
the angle $\phi=\phi(s_1^k,s_2^j)$ and $r_{12}=r_{12}(s_1^k,s_2^j)$ at
each of the centroids of the segments between the nodes
$k+1$ and $k$, where the length of the segments are assumed to be uniform and
denoted as $\Delta s_1$ for CNT (1) and $\Delta s_2$ for CNT (2).
As shown in Fig.~\ref{fig_projection}, the distance $r_{12}$
between the centroids of the segments $ds_1$ and $ds_2$ is obtained as
\begin{equation}
r_{12}= \bigg [(d_{1} - l_{x_{2}} + l_{x_{1}})^2
+ (l_{z_{1}} - l_{z_{2}})^2 \bigg]^{1/2} \;,
\label{eq:damp34}
\end{equation}
where $d_1$ is the spacing between the CNTs at the cathode substrate,
$l_{x_{1}}$ and $l_{x_{2}}$ are the deflections along $X$-axis,
and $l_{z_{1}}$ and $l_{z_{2}}$ are the deflections along $Z$-axis.
The angle of projection $\phi$ is expressed as
\begin{equation}
\phi = \tan^{-1} \bigg(\frac {l_{z_{1}}-l_{z_{2}}}{d_{1} - l_{x_{2}}
+ l_{x_{1}}}\bigg) \;.
\label{eq:damp35}
\end{equation}
The deflections $l_{x_{1}}$, $l_{z_{1}}$, $l_{x_{2}}$,
and $l_{z_{2}}$ are defined as, respectively,
\begin{equation}
l_{x_{1}}= \int_{0}^{s_{1}} ds_{1}\ \sin \theta_{1}
\equiv \sum_{j} \Delta s_{1} \sin \theta_{1} ^{j} \;
\label{eq:damp36}
\end{equation}
\begin{equation}
l_{z_{1}}= \int_{0}^{s_{1}} ds_{1}\ \cos \theta_{1}
\equiv \sum_{j} \Delta s_{1} \cos \theta_{1} ^{j} \;
\label{eq:damp37}
\end{equation}
\begin{equation}
l_{x_{2}}= \int_{0}^{s_{2}} ds_{2}\ \sin \theta_{2}
\equiv \sum_{j} \Delta s_{2} \sin \theta_{2} ^{j} \;
\label{eq:damp38}
\end{equation}
\begin{equation}
l_{z_{2}}= \int_{0}^{s_{2}} ds_{2}\ \cos \theta_{2}
\equiv \sum_{j} \Delta s_{2} \cos \theta_{2} ^{j} \;.
\label{eq:damp39}
\end{equation}
Note that the total electrostatic force on a particular CNT is to be
obtained by summing up all the binary contributions within the cell, that
is by summing up Eqs.~(\ref{eq:damp32}) and (\ref{eq:damp33}) over
the upper integer number of the quantity $N-1$, where $N$ is the number of
CNTs in the cell as discussed in Sec.~\ref{sec:evolution}.

\subsubsection{The van der Waals force}
Next, we consider the van der Waals effect. The van der Waals force plays important role not
only in the interaction of the CNTs with the substrate, but also in the interaction between
the walls of MWNTs and CNT bundles. Due to the overall effect of forces and flexibility of
the CNTs (here assumed to be elastic 1D members), the cylindrical symmetry of CNTs is
destroyed, leading to their axial and radial deformations. The change in cylindrical symmetry
may significantly affect the the properties of CNTs$^{37-38}$ 
. Here we
estimate the van der Waals forces due to the interaction between two concentric walls
of the MWCNTs.

Let us assume that the lateral and the longitudinal displacements of a CNT be $u_{x'}$ and
$u_{z'}$, respectively. We use updated Lagrangian approach with local coordinate system for
this description (similar to ($X',Z'$) system shown in Fig.~\ref{fig_electric}), where the
longitudinal axis coincides with $Z'$ and the lateral axis coincides with $X'$. Such a
description is consistent with the incremental procedure to update the CNT orientations in
the cells as adopted in the computational scheme. Also, due to the large length-to-diameter
ratio ($L(t)/d_t$), let the kinematics of the CNTs, which are idealized in this work as 1D
elastic members, be governed by that of an Euler-Bernoulli beam. Therefore, the kinematics
can be written as
\begin{equation}
u_{z'}^{(m)}= u_{z'0}^{(m)}- r^{(m)}
\frac {\partial u_{x'}^{(m)}}{\partial z'}\  \;,
\label{eq:damp40}
\end{equation}
where the superscript $(m)$ indicates the $m$th wall of the MWNT with
$r^{(m)}$ as its radius and $u_{z'0}$ is the longitudinal displacement of the
center of the cylindrical cross-section.
Under tension, bending moment and lateral shear force,
the elongation of one wall relative to its neighboring wall is
\begin{equation}
\Delta_{z'}^{(m)}= u_{z'}^{(m+1)} - u_{z'}^{(m)}
= r^{(m+1)}\frac{\partial u_{x'}^{(m+1)}}{\partial z'}
- r^{(m)}\frac{\partial u_{x'}^{(m)}}{\partial z'}
\approx (r^{(m+1)}- r^{(m)}) \frac{\partial \Delta_{x'}}{\partial s} \;,
\label{eq:damp41}
\end{equation}
where we assume $u_{x'}^{(m)}=u_{x'}^{(m+1)}=\Delta_{x'}$ as the lateral displacement as some
function of tensile force or compression buckling or pressure in the thin film device. The
lateral shear stress ($\tau_{vs}^{(m)}$) due to the van der Waals effect can now be written
as
\begin{equation}
\tau_{vs}^{(m)} = C_{vs} \frac {\Delta_{z}^{(m)}}{\Delta_{x}} \;,
\label{eq:damp42}
\end{equation}
where $\rm C_{vs}$ is the van der Waals coefficient.
Hence, the shear force per unit length can be obtained by integrating
Eq.~(\ref{eq:damp42}) over the individual wall circumferences and then
by summing up for all the neighboring pair interactions, that is,
\[
f_{vs}=\sum_m\int_{0}^{2 \pi}C_{vs}
\frac{\Delta_{z'}^{(m)}}{\Delta_{x'}} r_{\rm{eff}}\ d \psi
=\sum_m\int_{0}^{2 \pi}C_{vs} \frac{(r^{(m+1)}- r^{(m)})
\frac{\partial \Delta_{x'}}{\partial s}}{\Delta_{x'}}
\left(\frac {r^{(m+1)}+ r^{(m)}}{2}\right) \ d\psi
\]
\begin{equation}
\Rightarrow f_{vs}=\sum_m \pi C_{vs}
[(r^{(m+1)})^2-({r^{(m)}})^2]\frac{1}{\Delta_{x'}}
\frac{\partial \Delta_{x'}}{\partial s} \;.
\label{eq:damp44}
\end{equation}
The components of van der Waals force in the cell coordinate system ($X',Z'$) is then
obtained as
\begin{equation}
f_{vs_{z}}= f_{vs} \sin \theta(t) \;, \quad
f_{vs_{x}}=f_{vs} \cos \theta(t) \;.
\label{eq:damp45}
\end{equation}

\subsubsection{Ponderomotive force}
Ponderomotive force, which acts on free charges on the surface of CNTs, tends to straighten
the bent CNTs under the influence of electric field in the $Z$-direction. Furthermore, the
ponderomotive forces induced by the applied electric field stretch every
CNT$^{39}$ 
. We add this effect by assuming that the free charge at the tip
region is subjected to Ponderomotive force, which is computed as$^{40}$ 
\begin{equation}
f_{p_{z}}= \frac {1}{2N} \epsilon_{0} E_0^{2} \Delta A \cos \theta(t) \;,
\quad f_{p_{x}}= 0 \;,
\label{eq:damp46}
\end{equation}
where $\Delta A$ is the surface area of the cell on the anode side,
$f_{p_{z}}$ is the $Z$ component of the Ponderomotive force
and the $X$ component $f_{p_{x}}$ is assumed to be negligible.

\subsection{Modelling the reorientation of CNTs}
\label{sec:update}
The net force components acting on the CNTs along $Z$ and $X$
directions can be expressed as, respectively,
\begin{equation}
f_{z}= \int \left(f_{lz}+f_{vz_{z}}\right) ds + f_{c_{z}} + f_{p_{z}} \;,
\label{eq:damp47}
\end{equation}
\begin{equation}
f_{x}= \int \left(f_{lx}+f_{vs_{x}}\right) ds + f_{c_{x}} + f_{p_{x}} \;.
\label{eq:damp48}
\end{equation}
For numerical computation, at each time step the force components
obtained using Eqs.~(\ref{eq:damp47}) and (\ref{eq:damp48}) are
employed to update the curved shape $S'(x'+u_{x'},z'+u_{z'})$, where
the displacements are approximated using simple beam mechanics solution:
\begin{equation}
u_{z'} \approx \frac{1}{E'A_0}(f_{z}^{j+1} - f_{z}^{j})(z'^{j+1}-z'^j) \;,
\label{eq:damp49}
\end{equation}
\begin{equation}
u_{x'} \approx \frac{1}{3E'A_2} (f_{x}^{j+1}-f_{x}^j)
\left(x'^{j+1}-x'^j\right)^3 \;,
\label{eq:damp50}
\end{equation}
where $A_0$ is the effective cross-sectional area, $A_2$ is the area moment, $E'$ is the
modulus of elasticity for the CNT under consideration. The angle of orientation, $\theta(t)$,
of the corresponding segment of the CNT, that is between the node $j+1$ and node $j$, is
given by
\begin{equation}
\theta(t) =  \theta(t)^j =
\tan^{-1} \left(\frac{(x^{j+1}+u_{x}^{j+1})-(x^j+u_{x}^j)}
{(z^{j+1}+u_{z}^{j+1})-(z^j+u_{z}^j)}\right) \;,
\label{eq:damp51}
\end{equation}
\begin{equation}
\left\{ \begin{array}{c}
u_{x}^{j} \\ u_{z}^j
\end{array} \right\}=
\left[ \Gamma(\theta(t-\Delta t)^j)\right]
\left\{ \begin{array}{c}
u_{x'}^{j} \\ u_{z'}^j
\end{array} \right\} \;,
\end{equation}
where $\Gamma$ is the usual coordinate transformation matrix which
maps the displacements ($u_{x'},u_{z'}$) defined in the local ($X',Z'$)
coordinate system into the displacements ($u_x,u_z$) defined in the cell
coordinate system ($X,Z$). For this transformation, we employ the angle
$\theta(t-\Delta t)$ obtained in the previous time step and for each
node $j=1,2,\dots$.

\section{Computational scheme}
\label{sec:computation}
As already highlighted in the previous section, we model the CNTs as
generally oriented 1D elastic members. These 1D members are represented
by nodes and segments. With given initial distribution of the CNTs in the cell,
we discretize the time into uniform steps $t_{i+1}-t_i=\Delta t$.
The computational scheme involves three parts: (i) discretization of the
nucleation coupled model for degradation of CNTs derived in
Sec.~\ref{sec:evolution}, (ii) incremental update of the CNT geometry using
the estimated electromechanical force and (iii) computation
of the field emission current in the device.

\subsection{Discretization of the nucleation coupled model for degradation}
With the help of Eqs.~(\ref{eq:damp11})-(\ref{eq:damp13}) and by eliminating the kinetic
nucleation rate $N_{\rm{kin}}$, we first rewrite the simplified form of
Eqs.~(\ref{eq:damp7})-(\ref{eq:damp10}), which are given by, respectively,
\begin{equation}
S\frac{dn_1}{dt} - n_1\frac{dS}{dt} =
\frac{\beta_{11}n_s^2 S^3}{12}\sqrt{\frac{\Theta}{2\pi}}\exp\left[\Theta -
\frac{4\Theta^3}{27(\ln S)^2}\right] \;,
\label{eq:evol1}
\end{equation}
\begin{equation}
\frac{dS}{dt} = -\frac{2\beta_{11}n_s\Theta S}{81\sqrt{2\pi}(\ln S)^3}
\exp\left[\Theta - \frac{4\Theta^3}{27(\ln s)^2}\right]
- \sqrt{\frac{kT}{2\pi m_1}} (S-1) A_n \;,
\label{eq:evol2}
\end{equation}
\begin{equation}
\frac{dM_1}{dt} = \frac{\beta_{11}n_s^2 d_p^* S}{12}
\sqrt{\frac{\Theta}{2\pi}}\exp\left[\Theta-\frac{4\Theta^3}{27(\ln S)^2}\right]
+2n_s^2v_1\exp(\Theta)\sqrt{\frac{kT}{2\pi m_1}}(S-1) \;,
\label{eq:evol3}
\end{equation}
\begin{equation}
\frac{dA_n}{dt} = \frac{\beta_{11}n_s^2 s_1\Theta^{5/2}S}
{27\sqrt{2\pi}(\ln S)^2} \exp\left[\Theta -
\frac{4\Theta^3}{27(\ln S)^2}\right] +
4\pi n_s v_1 \sqrt{\frac{kT}{2\pi m_1}} M_1 (S-1) \;.
\label{eq:evol4}
\end{equation}
By eliminating $dS/dt$ from Eq.~(\ref{eq:evol1}) with the help of Eq.~(\ref{eq:evol2}) and by
applying a finite difference formula in time, we get
\[
\frac{n_{1_{i}}-n_{1_{i-1}}}{t_i-t_{i-1}} \approx
\frac{\beta_{11} n_{1_{i}}^2}{12}  \sqrt{\frac{\Theta}{2\pi}}
\exp{\left(- \frac{4\Theta^3}{27(\ln{S_{i-1}})^2}\right)} -
\frac{2\beta_{11}}{81} \frac {\Theta^{7/2}}{\sqrt{2\pi}}
\frac{n_{1_i}^2}{S_i}
\]
\begin{equation}
\frac {\exp {\bigg (\Theta- \frac{4\Theta ^ {3}}{27
(\ln{S_{i-1}})^{2}}\bigg)}}{(\ln{S_{i-1}})^{3}}+ 
\frac{n_{1_{i}}^2(S_{i}-1)A_{n}(i)}{S_{i}^2} \sqrt \frac {kT}{2\pi m_{1}} \;.
\label{eq:damp52}
\end{equation}
Similarly, Eqs.~(\ref{eq:evol2})-(\ref{eq:evol4}) are discretized as,
respectively,
\begin{equation}
\frac{S_i-S_{i-1}}{t_i-t_{i-1}} \approx
- \frac{2\beta_{11}}{81} \frac {\Theta^{7/2}}{\sqrt{2\pi}}
n_{1_{i}} \frac{\exp{\left(\Theta-
\frac{4\Theta^3}{27(\ln{S_{i-1}})^{2}}\right)}}{(\ln{S_{i-1}})^{3}}
- \frac{n_{1_{i}}(S_{i}-1)A_{n_{i}}}{S_{i}} \sqrt{\frac{kT}{2\pi m_{1}}}  \;,
\label{eq:damp53}
\end{equation}
\[
\frac{M_{1_{i}}-M_{1_{i-1}}}{t_i-t_{i-1}} \approx
\frac{\beta_{11}n_{1_{i}}^2}{12S_i}d_{p}^{*} \sqrt{\frac{\Theta}{2\pi}}
\exp{\left(\Theta- \frac{4\Theta^3}{27(\ln{S_{i-1}})^{2}}\right)}
\]
\begin{equation}
+ 2 v_{1} \frac{n_{1_i}^2 (S_i-1)}{S_i^2}\exp(\Theta)
\sqrt{\frac{kT}{2\pi m_{1}}}  \;,
\label{eq:damp531}
\end{equation}
\begin{equation}
\frac {A_{n_{i}}-A_{n_{i-1}}}{t_i-t_{i-1}} \approx
\frac{\beta_{11}s_1\Theta^{5/2}n_{1_{i}}}{27\sqrt{2\pi}}
\frac{\exp{\left(\Theta-\frac{4\Theta^3}{27(\ln{S_{i-1}})^{2}}\right)}}
{(\ln{S_{i-1}})^{2}}
+ 4 \pi v_{1}\sqrt{\frac{kT}{2\pi m_{1}}} (S_i-1)M_{1_{i}}  \;.
\label{eq:damp54}
\end{equation}
By simplifying Eq.~(\ref{eq:damp52}) with the help of
Eqs.~(\ref{eq:damp53})-(\ref{eq:damp54}), we get a quadratic polynomial of
the form
\begin{equation}
(b_1-b_2-b_3){n_{1_i}}^2-n_{1_{i}}+n_{1_{i-1}} = 0  \;,
\label{eq:damp62}
\end{equation}
where
\begin{equation}
b_{1}= \Delta t \frac{\beta_{11}}{12}
\sqrt{\frac{\Theta}{2\pi}}\exp{\left(- \frac{4\Theta ^ {3}}{27
(\ln{S_{i-1}})^{2}}\right)}  \;,
\label{eq:damp59}
\end{equation}
\begin{equation}
b_{2}= \Delta t \frac{2\beta_{11}}{81}
\frac{\Theta^{7/2}}{\sqrt{2\pi}}
\frac{\exp{\left(\Theta-
\frac{4\Theta^3}{27(\ln{S_{i-1}})^{2}}\right)}}{S_i(\ln{S_{i-1}})^3} \;,
\label{eq:damp60}
\end{equation}
\begin{equation}
b_{3}= \Delta t \frac{S_i-1}{S_i^2}A_{n_{i}}
\sqrt{\frac {kT}{2\pi m_{1}}}  \;.
\label{eq:damp61}
\end{equation}
Solution of Eq.~(\ref{eq:damp62}) yields two roots
(denoted by superscripts $(1,2)$):
\begin{equation}
n_{1_{i}}^{(1,2)} = \frac{1}{2(b_1-b_2-b_3)} \pm
\frac{\sqrt{1-4n_{1_{i-1}}(b_1-b_2-b_3)}}{2(b_1-b_2-b_3)}  \;.
\label{eq:damp63}
\end{equation}

For the first time step, the values of $b_{1}$, $b_{2}$ and $b_{3}$
are obtained by applying the initial conditions:
$S(0)=S_0$, $n_{1_{0}}=n_0$, and $A_{n_{0}}=A_{n0}$.
Since the $n_{1_{i}}$ must be real and finite, the following two
conditions are imposed:
$1-4n_{1_{i-1}}(b_1-b_2-b_3)\geq 0$ and $(b_{1}-b_{2}-b_{3}) \neq 0$.
Also, it has been assumed that the degradation of CNTs is an
irreversible process, that is, the reformation of CNTs
from the carbon cluster does not take place. Therefore, an additional
condition of positivity, that is, $n_{1_i} > n_{1_{i-1}}$ is introduced
while performing the time stepping.
Along with the above constraints, the $n_{1}$ history in a cell is
calculated as follows:
\begin{itemize}
\item
If $n_{1_i}^{(1)} > n_{1_{i-1}}$ and $n_{1_i}^{(1)} < n_{1_i}^{(2)}$,
then $n_{1_i} = n_{1_i}^{(1)}$;

\item
Else if $n_{1_{i}}^{(2)} > n_{1_{i-1}}$, then $n_{1_{i}} = n_{1_{i}}^{(2)}$;

\item
Otherwise the value of $n_{1}$ remains the same as in the previous time step,
that is, $n_{1_{i}} = n_{1_{i-1}}$.
\end{itemize}

Simplification of Eq.~(\ref{eq:damp53}) results in the following equation:
\begin{equation}
{S_i}^2+(c_1+c_2-S_{i-1}) S_i - c_1 = 0  \;,
\label{eq:damp65}
\end{equation}
where
\begin{equation}
c_1= \Delta t n_{1_{i}} A_{n_{i}} \sqrt{\frac{kT}{2\pi m_{1}}} \;,
\label{eq:damp66}
\end{equation}
\begin{equation}
c_2= \Delta t \frac{2\beta_{11}}{81} \frac{\Theta^{7/2}}{\sqrt{2\pi}}
n_{1_{i}}
\frac{\exp{\left(\Theta-
\frac{4\Theta^3}{27(\ln{S_{i-1}})^{2}}\right)}}{(\ln{S_{i-1}})^{3}}  \;.
\label{eq:damp67}
\end{equation}
Solution of Eq.~(\ref{eq:damp65}) yields the following two roots:
\begin{equation}
S_i=-\frac{1}{2}(c_1+c_2-S_{i-1}) \pm
\frac{1}{2}\sqrt{c_1+c_2-S_{i-1}^2+4c_1}  \;.
\label{eq:damp68}
\end{equation}
For the first time step, $c_1$ and $c_2$ are calculated
with the following conditions:
$n_{1_{1}}$ from the above calculation, $S(0)=S_0$, and
$A_{n_{0}} = A_{n0}$.
Realistically, the saturation ratio $S$ cannot be negative
or equal to one. Therefore, $S_i>0$ yields
$c_1>0$. 

While solving for $A_{n}$, the Eq.~(\ref{eq:damp54}) is 
solved with the values of $n_{1}$ and
$S$ from the above calculations and the initial conditions $A_{n_0} = A_{n0}$,
$M_{1_{0}}=M_0$. The value of $M_{1_{0}}$ was calculated by assuming $n(d_{p},t)$ as a
standard normal distribution function.

\subsection{Incremental update of the CNT geometry}
At each time time step $t=t_i$, once the $n_{1_{i}}$ is solved,
we are in a position to compute the net electromechanical force (see
Sec.~\ref{sec:electromechanical}) as
\begin{equation}
f_i = f_i(E_0,n_{1_{i-1}},\theta(t_{i-1})) \;.
\end{equation}
Subsequently, the orientation angle for each segment of each CNT is
then obtained as (see Sec.~\ref{sec:update})
\begin{equation}
\theta(t_i)^j = \theta(f_i)^j \;
\end{equation}
and it is stored for future calculations. A critical angle, ($\theta_c$), is generally
employed with $\theta_c\approx \pi/4$ to $\pi/2.5$ for the present numerical simulations. For
$\theta\le\theta_c$, the meaning of $f_{z}$ is the ``longitudinal force'' and the meaning of
$f_{x}$ is the ``lateral force'' in the context of Eqs.~(\ref{eq:damp49}) and
(\ref{eq:damp50}). When $\theta>\theta_c$, the meanings of $f_{z}$ and $f_{x}$ are
interchanged.

\subsection{Computation of field emission current}
\begin{figure}
\centerline{\psfig{file=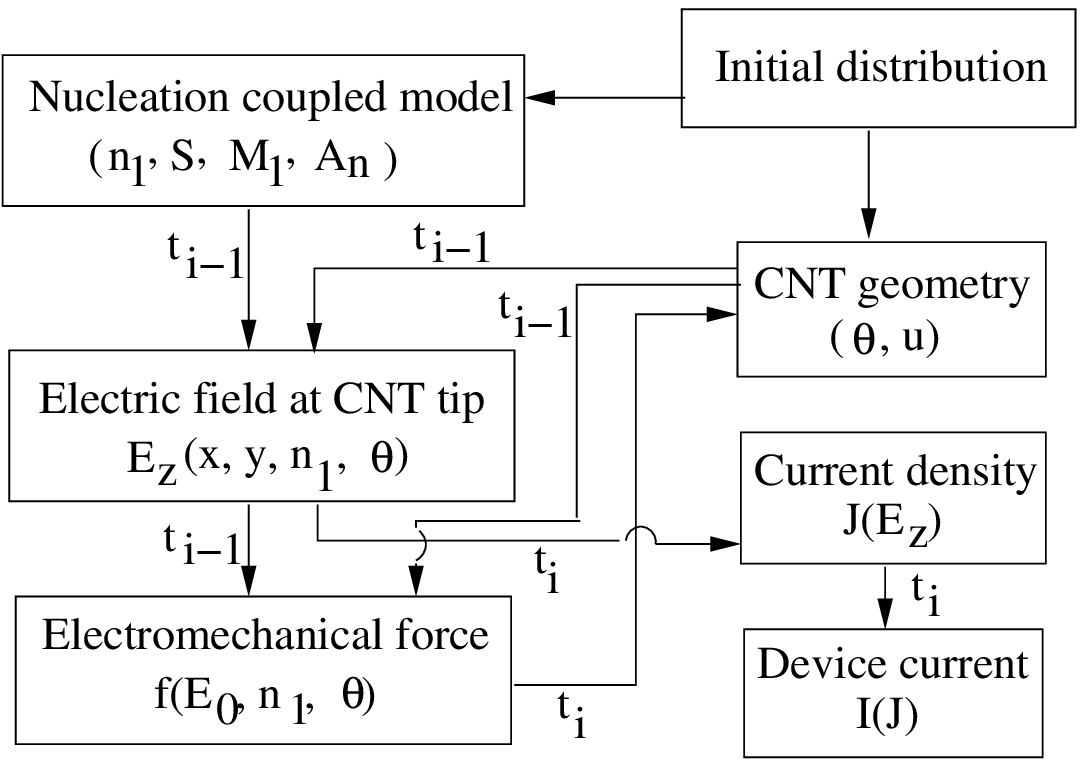,width=7cm}}
\caption{Computational flow chart for calculating the device current.}
\label{fig_computation}
\end{figure}

Once the updated tip angles and the electric field at the tip are obtained at a particular
time step, we employ Eq.~(\ref{eq:damp1}) to compute the current density contribution from
each CNT tip, which can be rewritten as
\begin{equation}
J_i = \frac{BE_{z_i}^2}{\Phi}\exp\left(-\frac{C\Phi^{3/2}}{E_{z_{i}}}\right)\;,
\end{equation}
with $B=(1.4\times 10^{-6})\times \exp(9.8929\times \Phi^{-1/2})$ and $C = 6.5\times 10^{7}$
taken from ref.$^{41}$ 
. The device current ($I_i$) from each computational cell
with surface area $\Delta A$ at the anode at the present time step $t_i$ is obtained by
summing up the current density over the number of CNTs in the cell, that is,
\begin{equation}
I_i = \Delta A \sum_{\approx N} J_i \;.
\end{equation}
Fig.~\ref{fig_computation} shows the flow chart of the computational scheme
discussed above.

At $t=0$, in our model, the CNTs can be randomly oriented. This random distribution is
parameterized in terms of the upper bound of the CNT tip deflection, which is given by
$\Delta x_{\max}=h/q$, where $h$ is the CNT length and $q$ is a real number. In the numerical
simulations which will be discussed next, the initial tip deflections can vary widely. The
following values of the upper bound of the tip deflection have been considered: $\Delta
x_{\max}=h_0/(5+10p)$, $(p=0,1,2, ... ,9)$. The tip deflection $\Delta x$ is randomized
between zero and these upper bounds. Simulation for each initial input with a randomized
distribution of tip deflections was run for a number of times and the maximum, minimum, and
average values of the output current were obtained. In the first set, the simulations were
run for a uniform height, radius and spacing of CNTs in the film. Subsequently, the height,
the radius and the spacing were varied randomly within certain bounds, and their effects on
the output current were analyzed.

\section{Results and discussions}
\label{sec:results} The CNT film under study in this work consists of randomly oriented
multi-walled nanotubes (MWNTs). The film samples were grown on a stainless steel substrate.
The film has a surface area of $1 cm^{2}$ and thickness of $10-14 \mu m$. The anode consists
of a $1.59 mm$ thick copper plate with an area of $49.93 mm^{2}$. The current-voltage history
is measured over a range of DC bias voltages for a controlled gap between the cathode and the
anode. In the experimental set-up, the device is placed within a vacuum chamber of a
multi-stage pump. The gap ($d$) between the cathode substrate and the anode is controlled
from outside by a micrometer.

\subsection{Degradation of the CNT thin films}
\begin{figure}
\centerline{\psfig{file=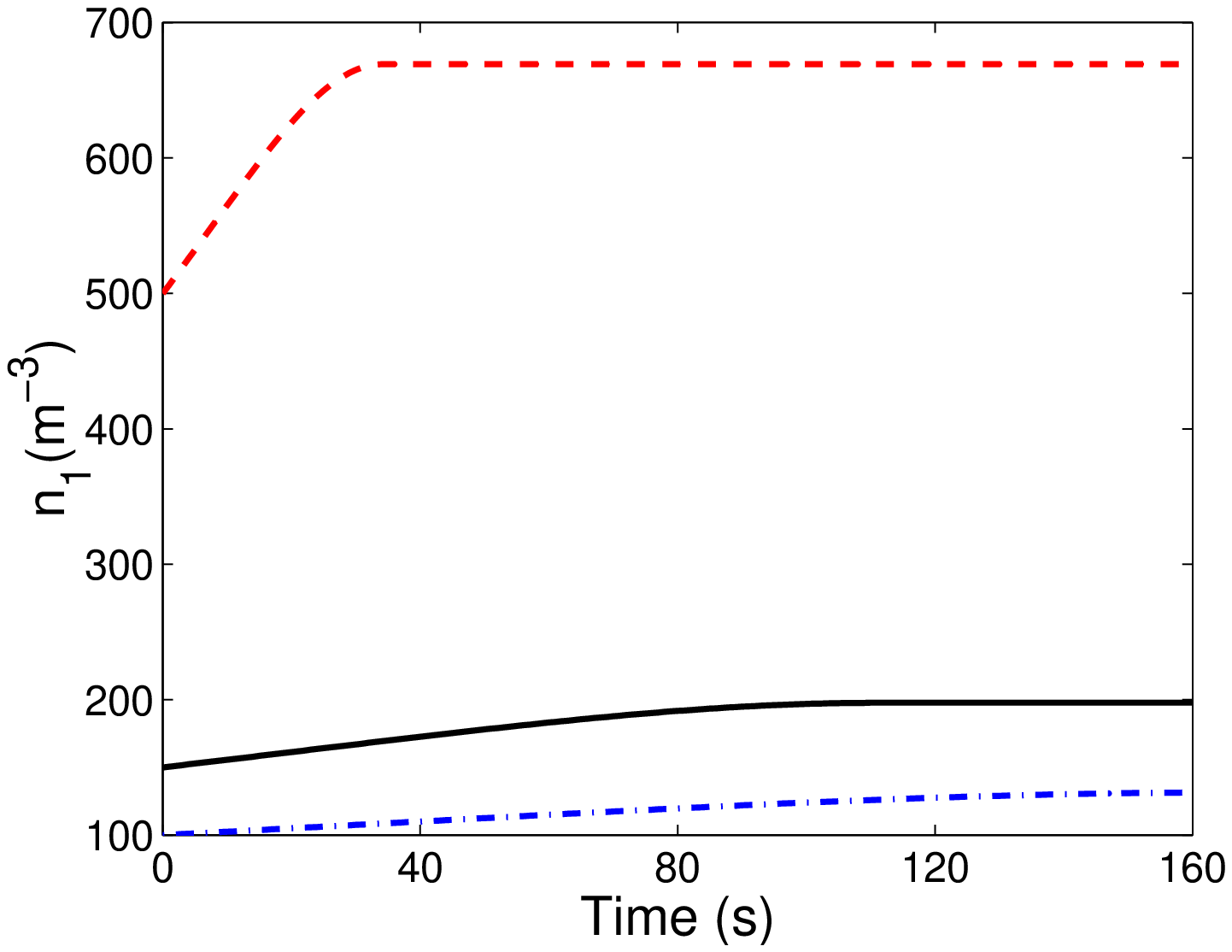,width=7cm}}
\caption{Variation of carbon cluster concentration over
time. Initial condition: $S(0)=100$, $T=303K$,
$M_{1}(0)= 2.12\times10^{-16}$, $A_n(0)=0$.}
\label{fig_n1}
\end{figure}

We assume that at $t=0$, the film contains negligible amount of carbon cluster. To understand
the phenomena of degradation and fragmentation of the CNTs, following three sets of input are
considered: $n_{1}(0)=100,\, 150,\, 500$. The other initial conditions are set as $S(0)=100$,
$M_{1}(0)= 2.12\times10^{-16}$, $A_n(0)=0$, and $T=303K$. Fig.~\ref{fig_n1} shows the three
$n_1(t)$ histories over a small time duration ($160 s$) for the three cases of $n_{1}(0)$,
respectively. For $n_{1}(0)=100$ and $150$, the time histories indicate that the rate of
decay is very slow, which in turn implies longer lifetime of the device. For $n_{1}(0)=500$,
the time history indicates that the CNTs decay comparatively faster, but still insignificant
for the first $34s$, and then the cluster concentration becomes constant. It can be
concluded from the above three cases that the rate of decay of CNTs is generally slow under
operating conditions, which implies stable performance and longer lifetime of the device if
this aspect is considered alone.

\begin{figure}
\centerline{\psfig{file=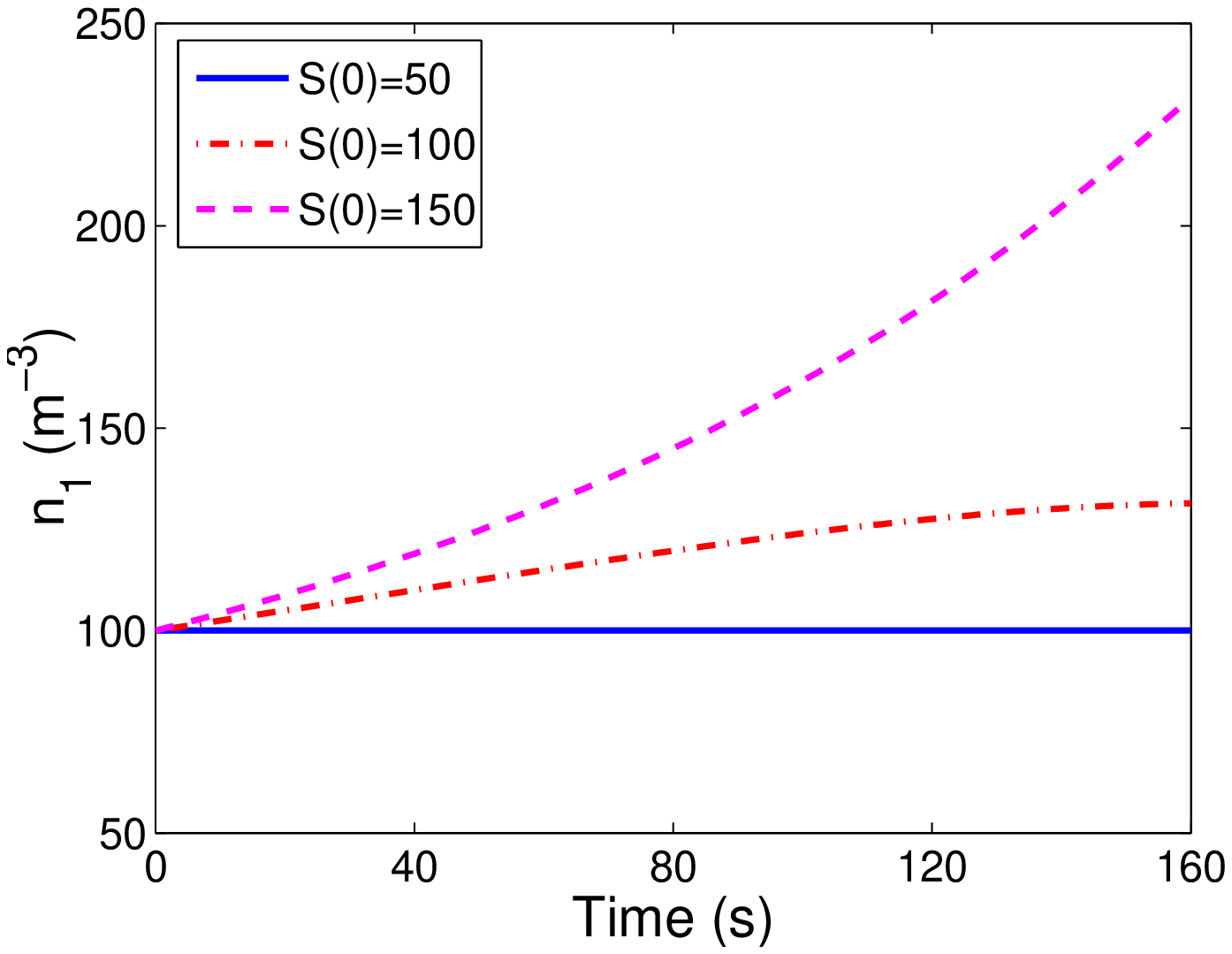,width=7cm}}
\caption{Variation of carbon cluster concentration over
time. Initial condition: $n_1(0)=100 m^{-3}$, $T=303K$,
$M_{1}(0)= 2.12\times10^{-16}$, $A_n(0)=0$.}
\label{fig_S}
\end{figure}

Next, the effect of variation in the initial saturation ratio $S(0)$ on $n_{1}(t)$ history is
studied. The value of $n_{1}(0)$ is set as $100$, while other parameters are assumed to have
identical value as considered previously. The following three initial conditions in $S(0)$
are considered: $S(0)=50,\, 100,\, 150$. Fig.~\ref{fig_S} shows the $n_1(t)$ histories. It
can be seen in this figure that for $S(0)=100$ (moderate value), the carbon cluster
concentration first increases and then tends to a steady state. This was also observed in
Fig.~(\ref{fig_n1}). For higher values of $S(0)$, $n_1$ increases exponentially over time.
For $S(0)=50$, a smaller value, the decay is not observed at all. This implies that a small
value of $S(0)$ is favorable for longer lifetime of the cathode. However, a more detailed
investigation on the physical mechanism of cluster formation and CNT fragmentation may be
necessary, which is an open area of research.

\begin{figure}
\centerline{\psfig{file=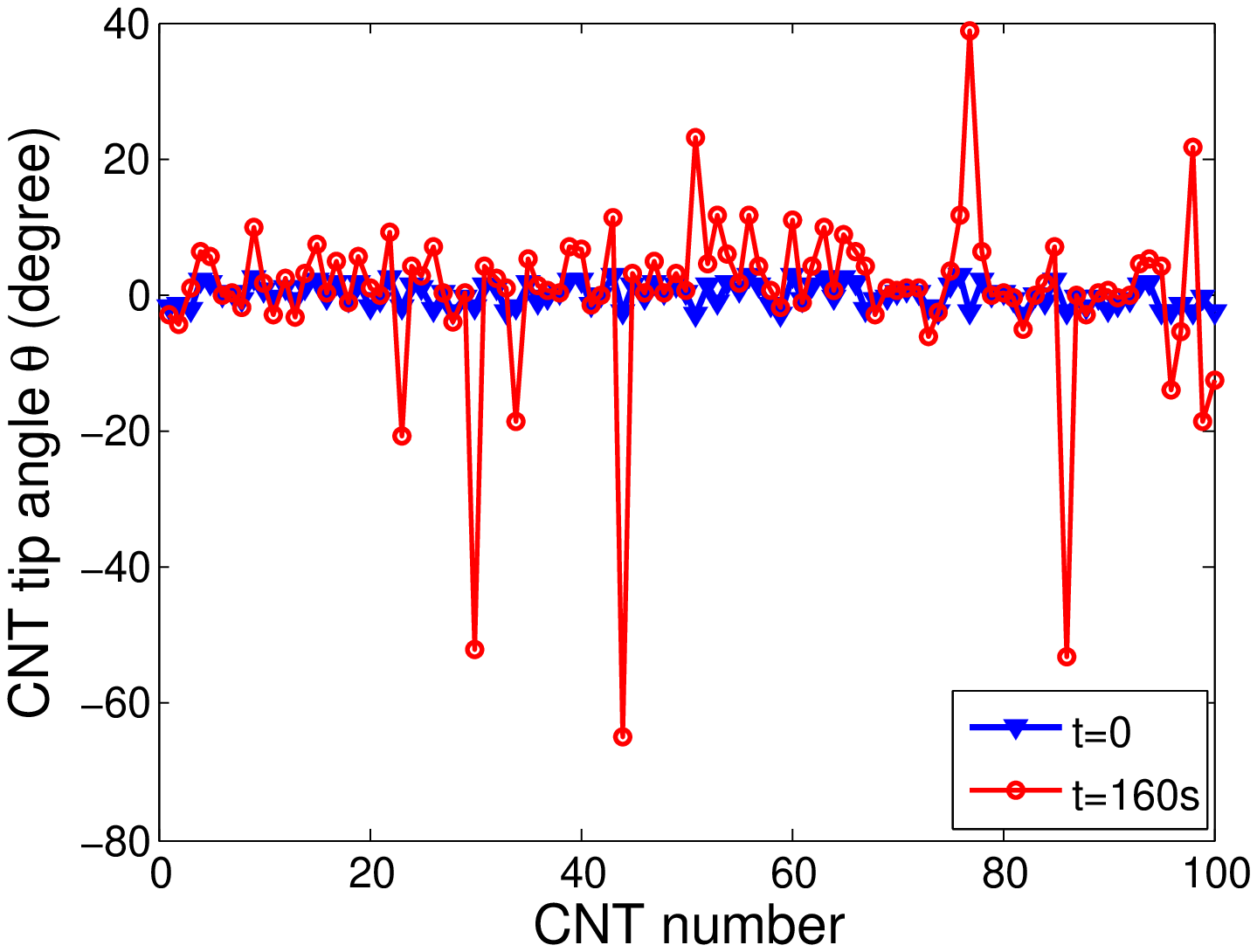,width=7cm}}
\caption{Distribution of tip angles over the number of CNTs.}
\label{fig_angle}
\end{figure}

At $t=0$, we assign random orientation angles ($\theta(0)^j$) to the CNT segments. For a cell
containing 100 CNTs, Fig.~\ref{fig_angle} shows the terminal distribution of the CNT tip
angles (at $t=160s$ corresponding to the $n_{1}(0)=100$ case discussed previously) compared
to the initial distribution (at $t=0$). The large fluctuations in the tip angles for many of
the CNTs can be attributed to the significant electromechanical interactions.

\subsection{Current-voltage characteristics}
\label{sec:IV} In the present study, the quantum-mechanical treatment has not been explicitly
carried out, and instead, the Fowler-Nordheim equation has been used to calculate the current
density. In such a semi-empirical calculation, the work function $\Phi$$^{42}$
for
the CNTs must be known accurately under a range of conditions for which the device-level
simulations are being carried out. For CNTs, the field emission electrons originate from
several excited energy states (non metallic electronic
states)$^{43-44}$ 
. Therefore, the the work function for CNTs is
usually not well identified and is more complicated to compute than for metals. Several
methodologies for calculating the work function for CNTs have been proposed in literature. On
the experimental side, Ultraviolet Photoelectron Spectroscopy (UPS) was used by Suzuki {\it
et al.}$^{45}$ 
to calculate the work function for SWNTs. They reported a work
function value of 4.8 eV for SWNTs. By using UPS, Ago {\it et al.}$^{46}$ 
measured
the work function for MWNTs as 4.3 eV. Fransen {\it et al.}$^{47}$ 
used the field
emission electronic energy distribution (FEED) to investigate the work function for an
individual MWNT that was mounted on a tungsten tip. Form their experiments, the work function
was found to be $7.3 \pm 0.5$ eV. Photoelectron emission (PEE) was used by Shiraishi {\it et
al.}$^{48}$ 
to measure the work function for SWNTs and MWNTs. They measured the
work function for SWNTs to be 5.05 eV and for MWNTs to be 4.95 eV. Experimental estimates of
work function for CNTs were carried out also by Sinitsyn {\it et al.}$^{49}$ 
.
Two types were investigated by them: (i) 0.8-1.1 nm diameter SWNTs twisted into ropes of 10
nm diameter, and (ii) 10 nm diameter MWNTs twisted into 30-100 nm diameter ropes. The work
functions for SWNTs and MWNTs were estimated to be 1.1 eV and 1.4 eV, respectively. Obraztsov
{\it et al.}$^{50}$ 
reported the work function for MWNTs grown by CVD to be in
the range 0.2-1.0 eV. These work function values are much smaller than the work function
values of metals ($\approx 3.6-5.4 eV$), silicon($\approx 3.30-4.30 eV$), and
graphite($\approx 4.6-5.4 eV$). The calculated values of work function of CNTs by different
techniques is summarized in Table 1. The wide range of work functions in different studies
indicates that there are possibly other important effects (such as electromechanical
interactions and strain) which also depend on the method of sample preparation and different
experimental techniques used in those studies. In the present study, we have chosen $\Phi=2.2
eV$.

\begin{figure}
\centerline{\psfig{file=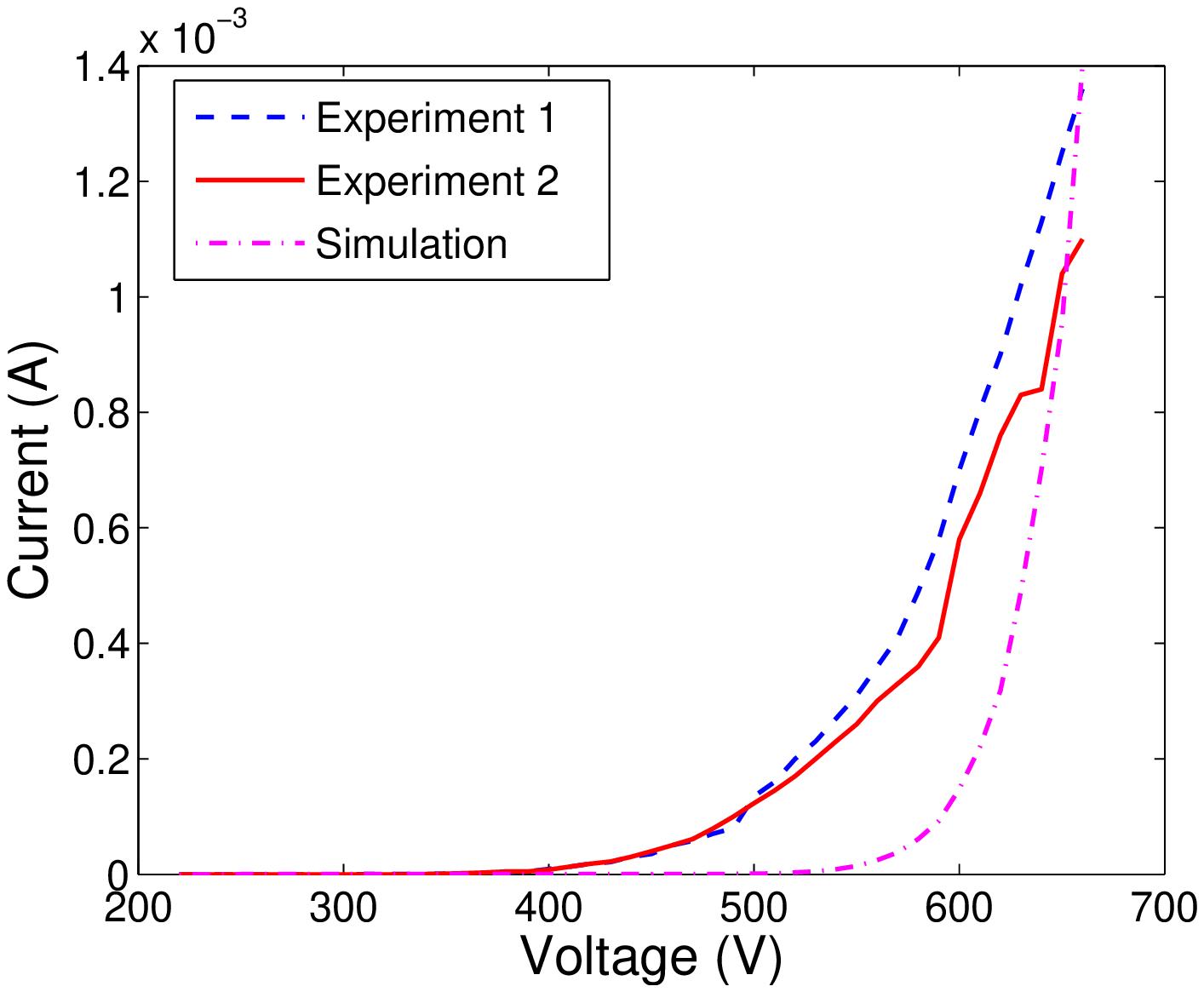,width=7cm}}
\caption{Comparison of simulated current-voltage characteristics with
experiments.}
\label{fig_IV}
\end{figure}

The simulated current-voltage (I-V) characteristics
of a film sample for a gap $d=34.7 \mu m$ is compared with the
experimental measurement in Fig.~\ref{fig_IV}.
The average height, the average radius and the average spacing between
neighboring CNTs in the film sample are taken as
$h_0=12 \mu m$, $r=2.75 nm$, and $d_1=2 \mu m$.
The simulated I-V curve in Fig.~\ref{fig_IV} corresponds to the
average of the computed current for the ten runs. This is the first and
preliminary simulation of its kind based on a multiphysics
based modeling approach and the present model predicts the I-V
characteristics which is in close agreement with the experimental measurement.
However, the above comparison indicates that there are some deviations near
the threshold voltage of $\approx 500-600 V$, which needs to be looked at
by improving the model as well as experimental materials and method.

\subsection{Field emission current history}
Next, we simulate the field emission current histories for the similar sample configuration
as used previously, but for three different parametric variations: height, radius, and
spacing. Current histories are shown for constant bias voltages of $440V$, $550V$ and $660V$.

\subsubsection{Effects of uniform height,
uniform radius and uniform spacing} 
\begin{figure}
\centerline{\psfig{file=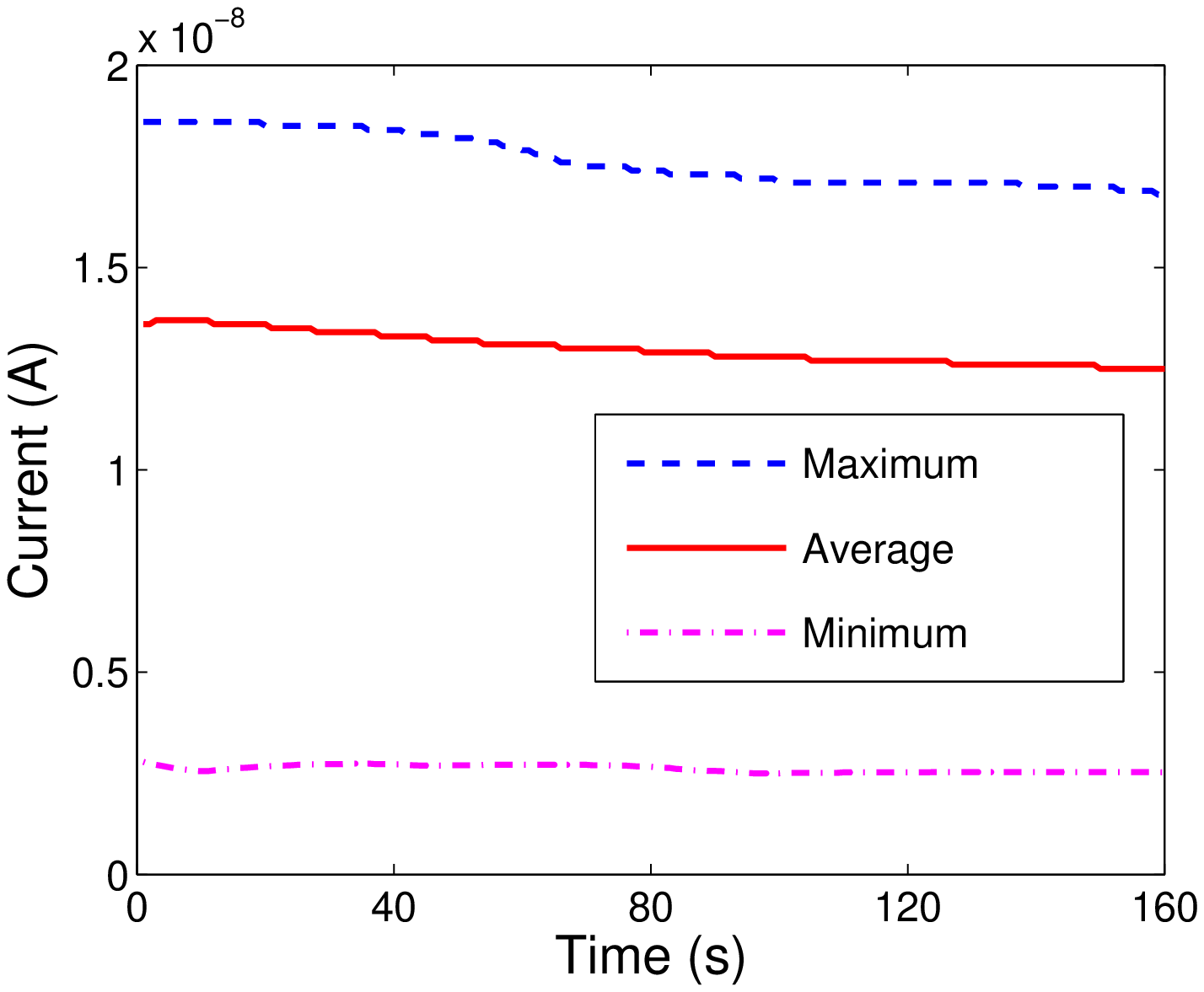,width=5.25cm}\hfill
            \psfig{file=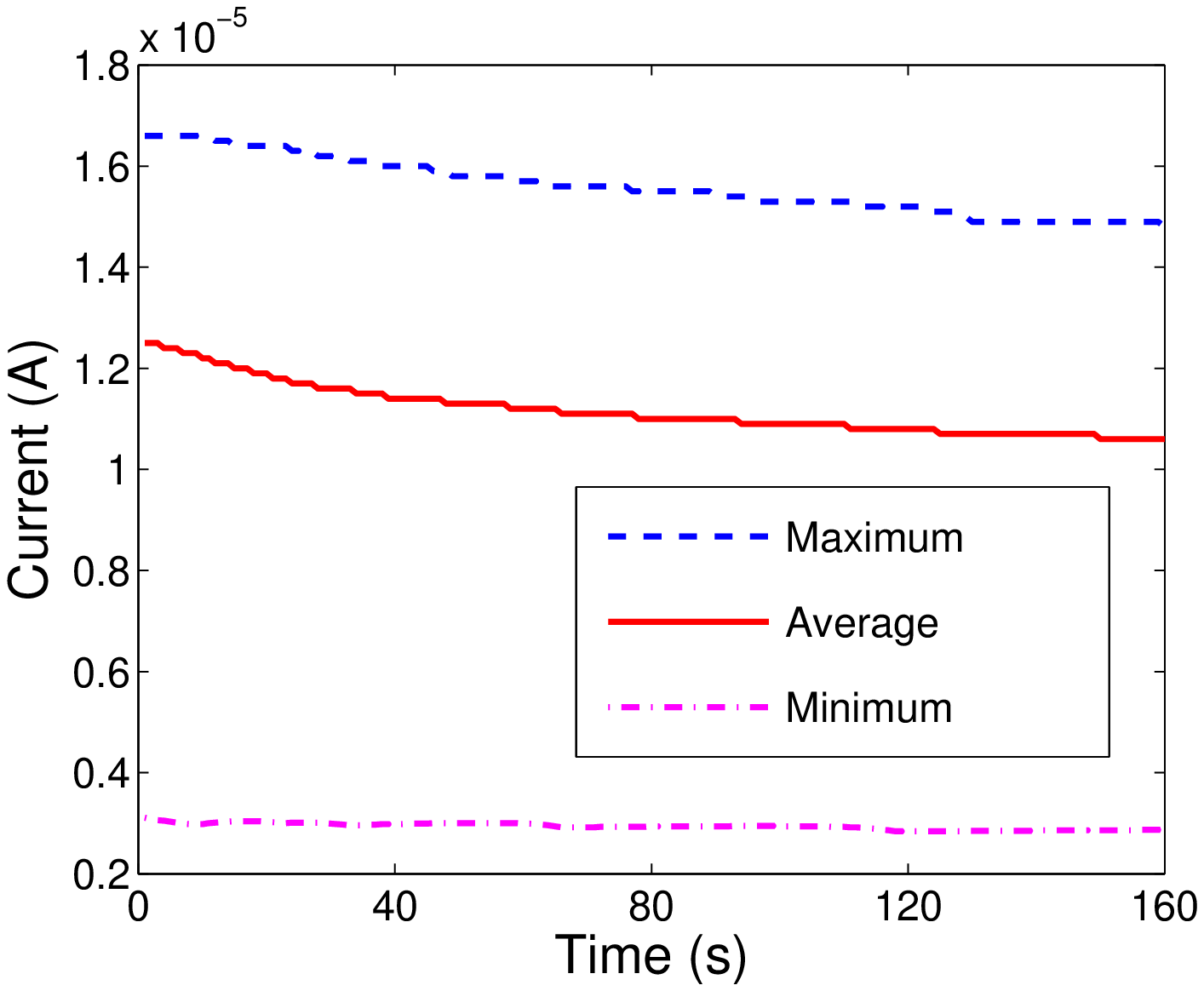,width=5.25cm}\hfill
            \psfig{file=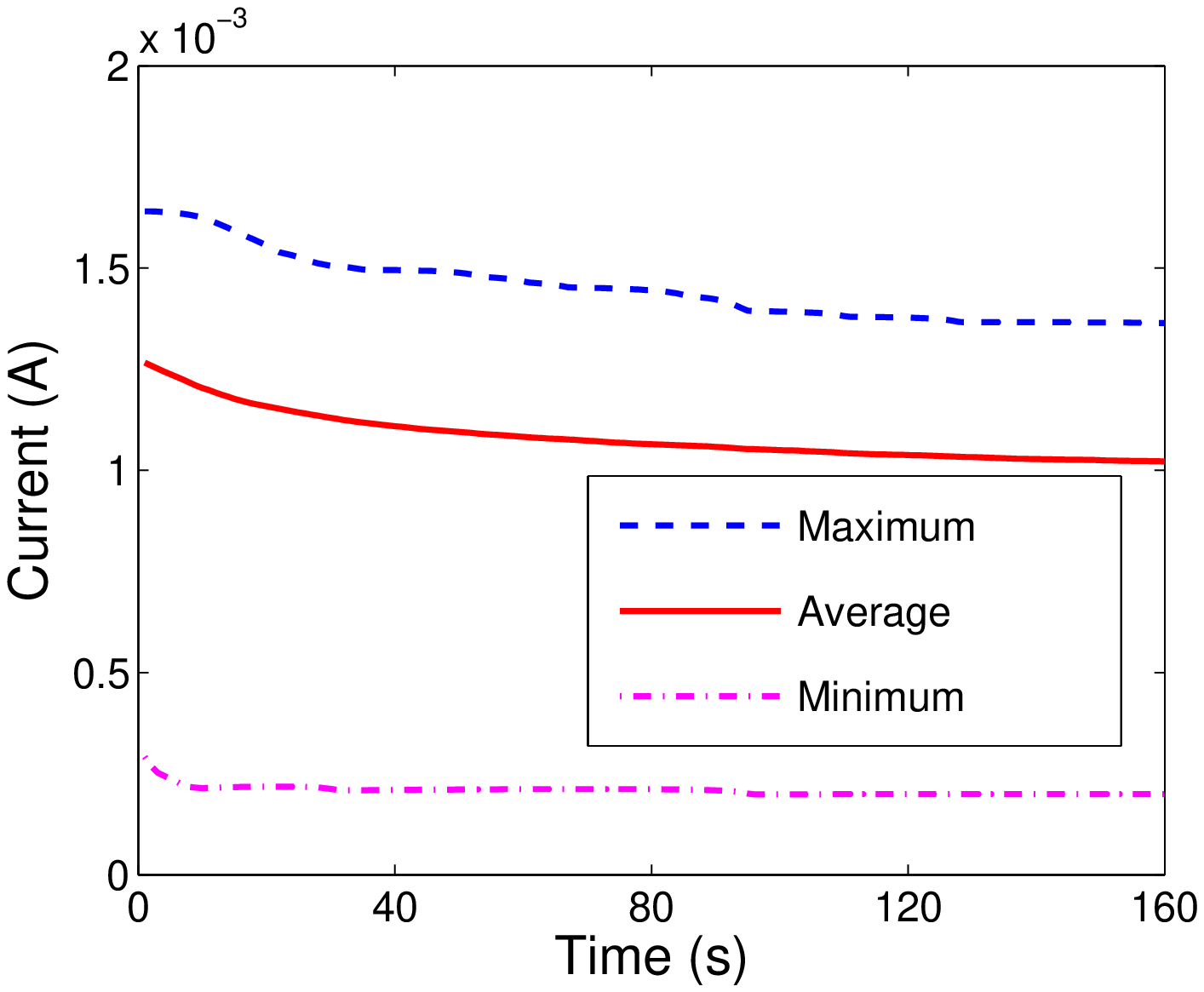,width=5.25cm}}
\centerline{(a)\hspace{5cm}(b)\hspace{5cm} (c)}
\caption{Simulated current histories for
uniform radius, uniform height and uniform spacing of CNTs at a bias
voltage of (a) 440 V, (b) 550 V, and (c) 660 V.}
\label{fig_uniform}
\end{figure}

In this case, the values of height, radius, and the
spacing between the neighboring CNTs are kept identical to the previous current-voltage
calculation in Sec.~\ref{sec:IV}. Fig.~\ref{fig_uniform}(a), (b) and (c) show the current
histories for three different bias voltages of $440V$, $550V$ and $660V$. In the subfigures,
we plot the minimum, the maximum and the average currents over time as post-processed from a
number of runs with randomized input distributions. At a bias voltage of $440 V$, the average
current decreases from $1.36 \times 10^{-8}A$  to $1.25 \times 10^{-8}A$ in steps. The
maximum current varies between $1.86 \times 10^{-8}A$  to $1.68 \times 10^{-8}A$, whereas the
minimum current varies between $2.78 \times 10^{-9}A$ to $2.52 \times 10^{-9}A$. Comparisons
among the scales in the sub-figures indicate that there is an increase in the order of
magnitude of current when the bias voltage is increased. The average current decreases from
$1.25 \times 10^{-5}A$ to $1.06 \times 10^{-5}A$ in steps when the bias voltage is increased
from $440V$ to $550 V$. At the bias voltage of $660 V$, the average value of the current
decreases from $1.26 \times 10^{-3}A$ to $1.02 \times 10^{-3}A$. The increase in the order of
magnitude in the current at higher bias voltage is due to the fact that the electrons are
extracted with a larger force. However, at a higher bias voltage, the current is found to
decay faster (see Fig.~\ref{fig_uniform}(c)).

\subsubsection{Effects of non-uniform radius}
\begin{figure}
\centerline{\psfig{file=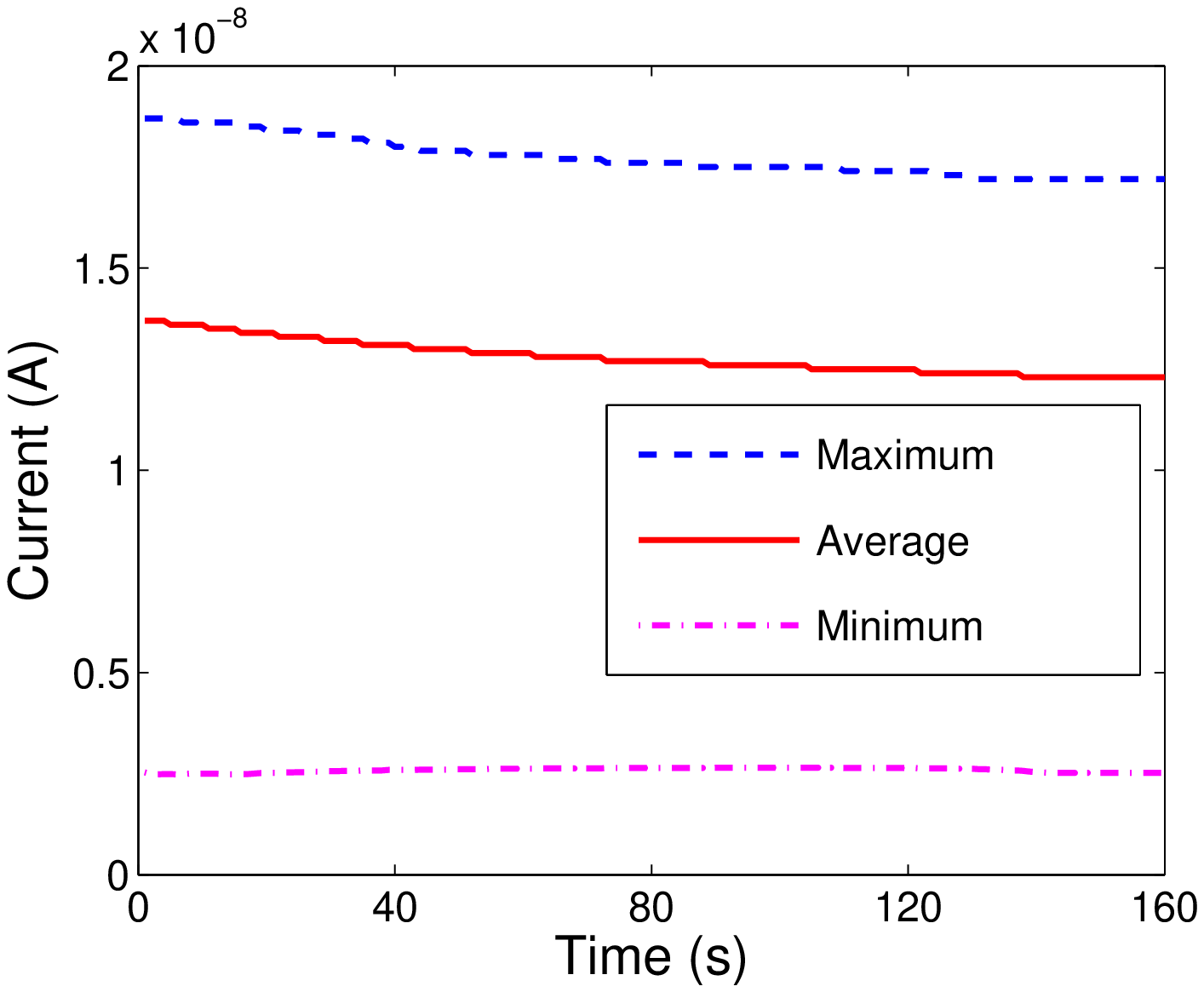,width=5.25cm}\hfill
            \psfig{file=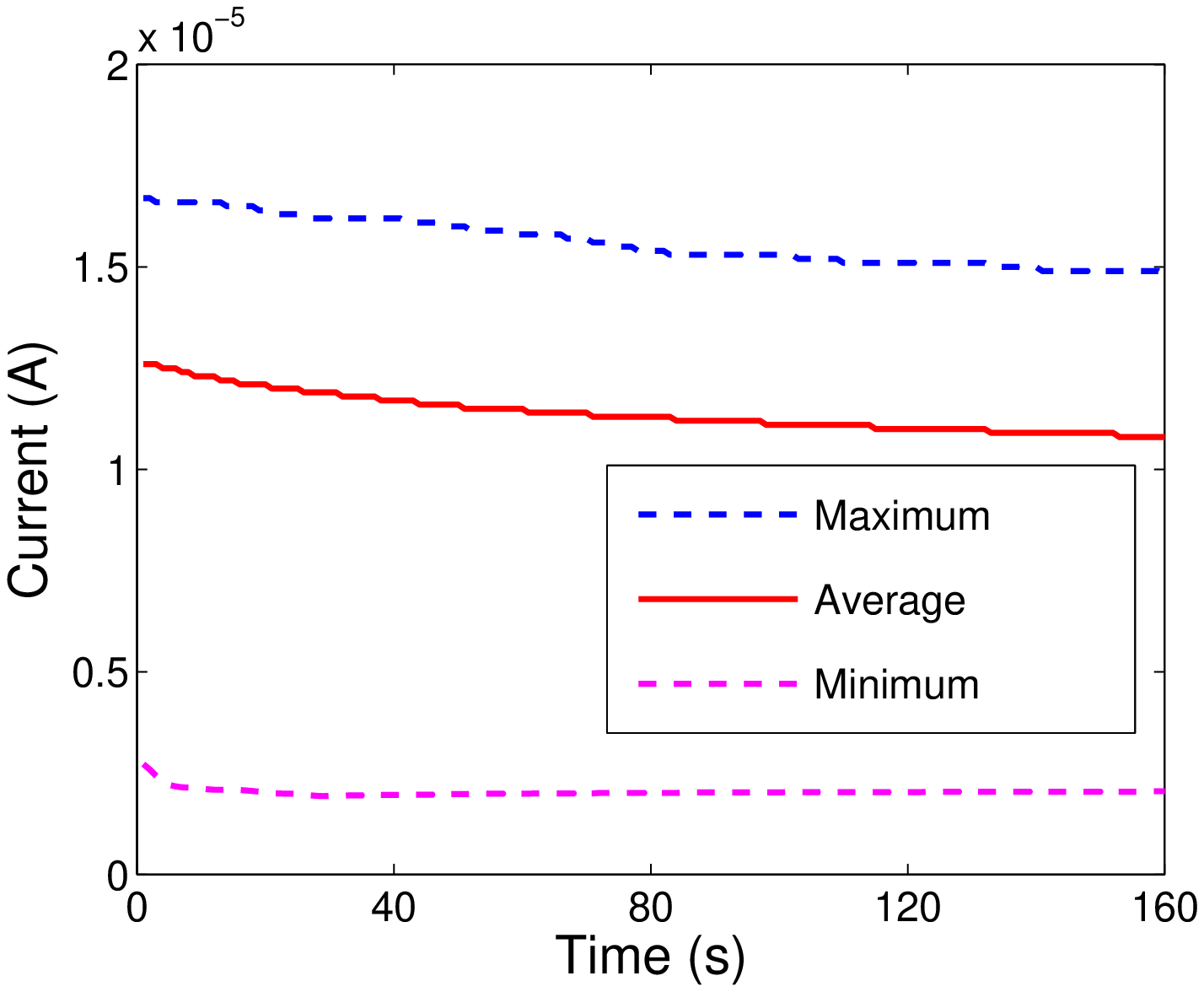,width=5.25cm}\hfill
            \psfig{file=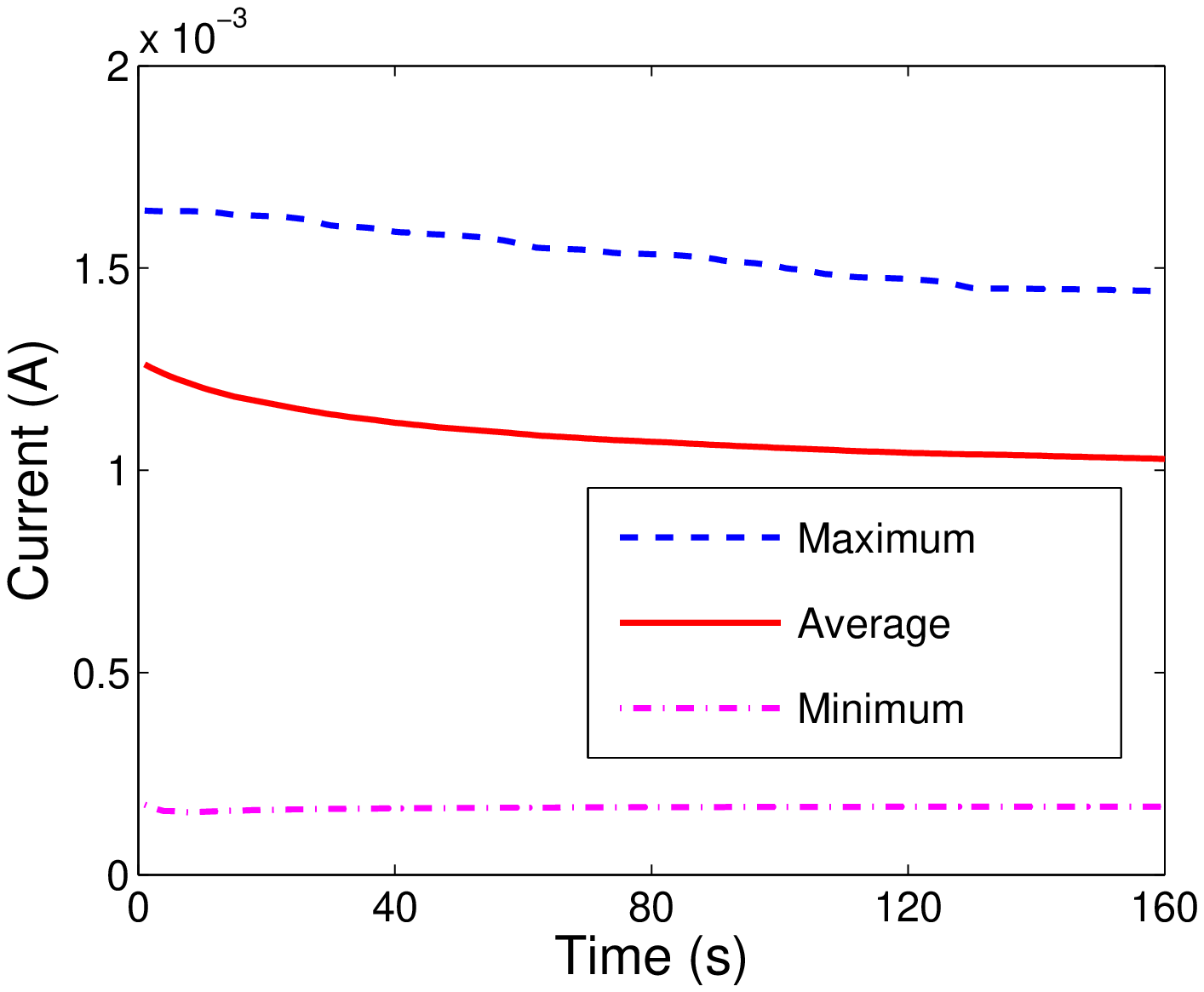,width=5.25cm}}
\centerline{(a)\hspace{5cm}(b)\hspace{5cm}(c)}
\caption{Simulated current histories for
non-uniform radius of CNTs at a bias
voltage of (a) 440 V, (b) 550 V, and (c) 660 V.}
\label{fig_radius}
\end{figure}

In this case, the uniform height and the uniform spacing between the neighboring CNTs are
taken as $h_0=12 \mu m$ and $d_1=2 \mu m$, respectively. Random distribution of radius is
given with bounds $1.5 - 4 nm$. The simulated results are shown in Fig.~\ref{fig_radius}. At
the bias voltage of $440 V$, the average current decreases from $1.37 \times 10^{-8}A$ at
$t=1s$ to $1.23 \times 10^{-8}A$ at $t=138s$ in steps and then the current stabilizes. The
maximum current varies between $1.87 \times 10^{-8}A$  to $1.72 \times 10^{-8}A$, whereas the
minimum current varies between $2.53 \times 10^{-9}A$ to $2.52 \times 10^{-9}A$. The average
current decreases from $1.26 \times 10^{-5}A$ to $1.08 \times 10^{-5}A$ in steps when the
bias voltage is increased from $440 V$ to $550 V$. At a bias voltage of $660 V$, the average
current decreases from $1.26 \times 10^{-3}A$ to $1.02 \times 10^{-3}A$. As expected, a more
fluctuation between the maximum and the minimum current have been observed here when compared
to the case of uniform radius.

\subsubsection{Effects of non-uniform height}
\begin{figure}
\centerline{\psfig{file=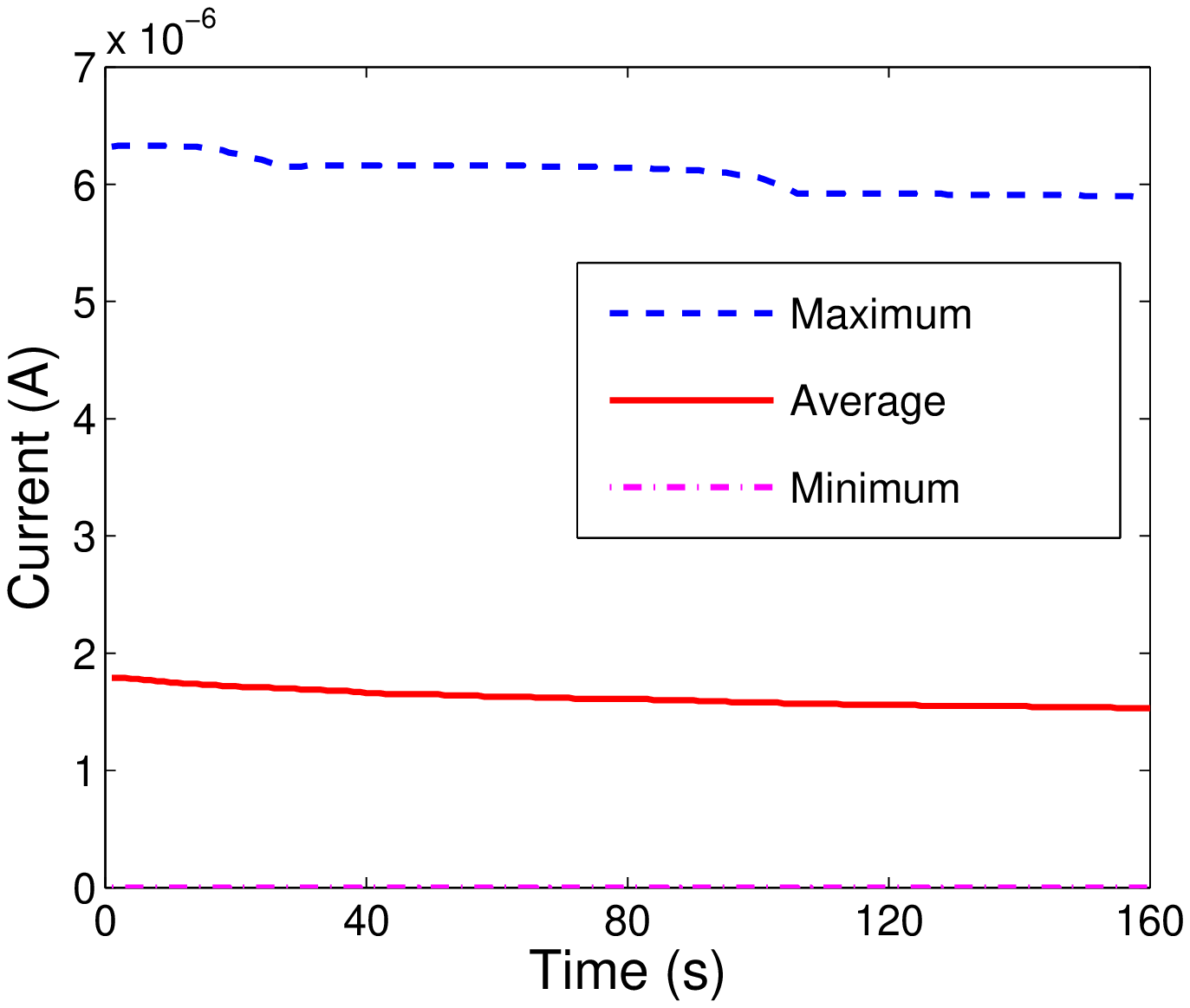,width=5.25cm}\hfill
            \psfig{file=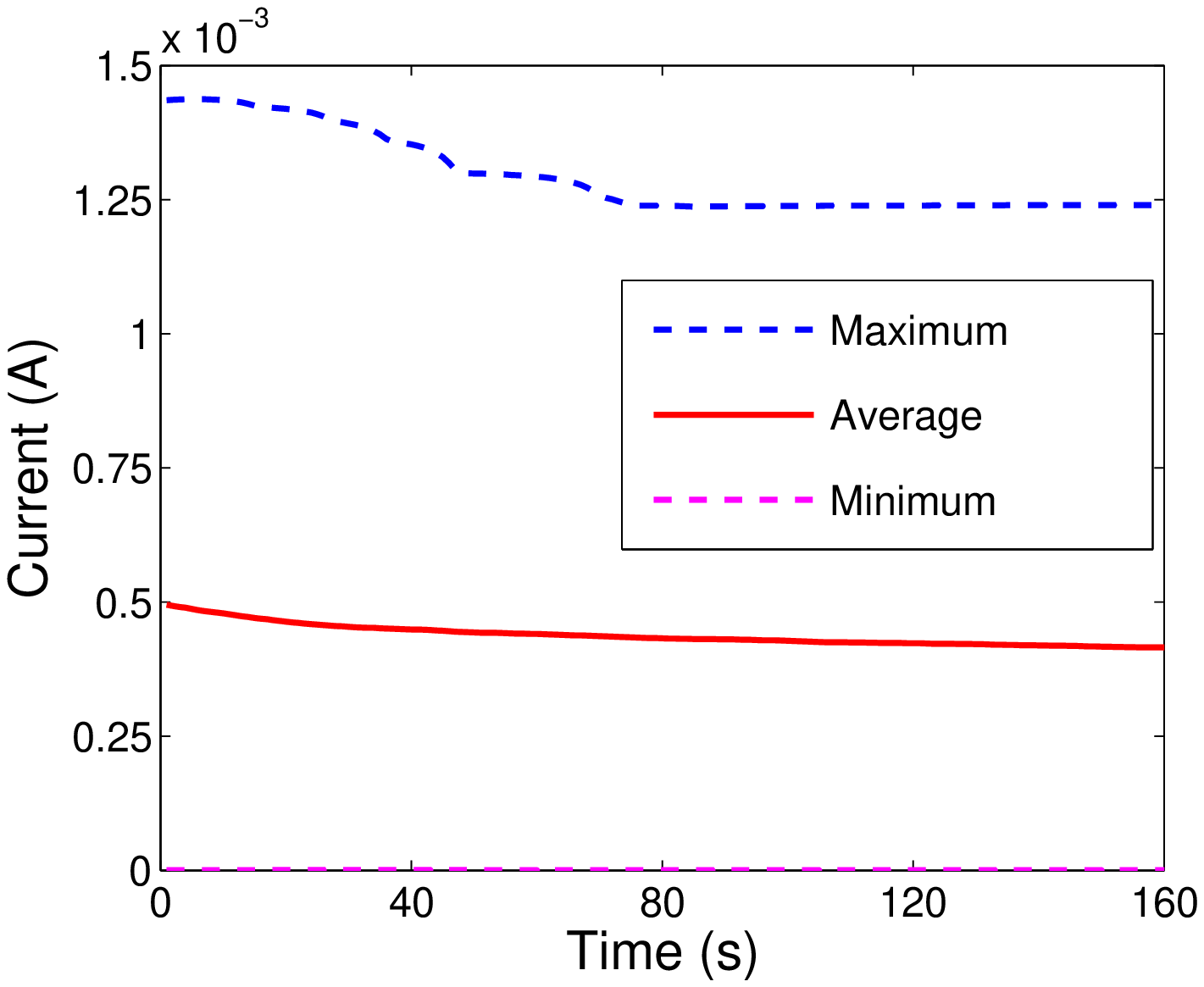,width=5.25cm}\hfill
            \psfig{file=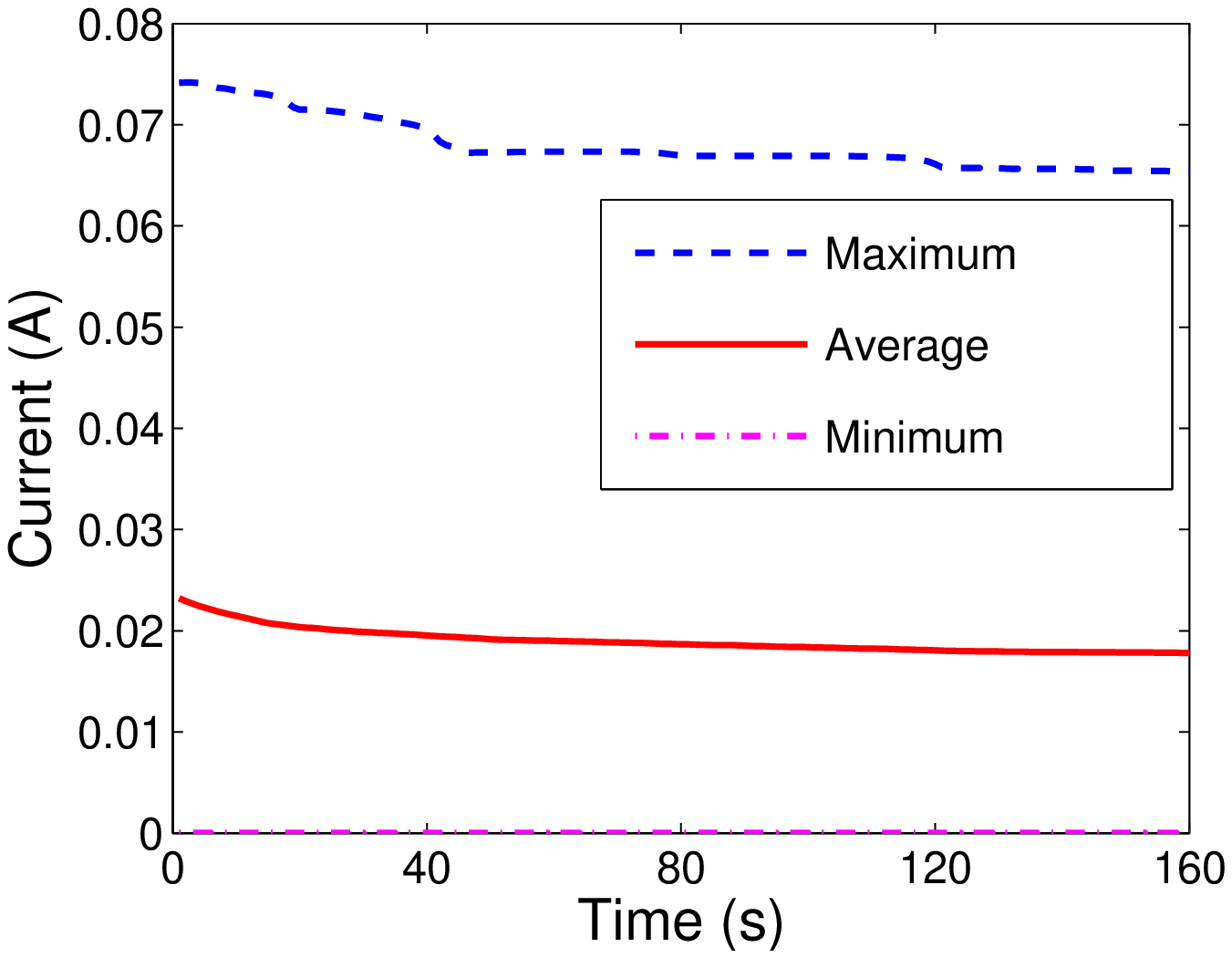,width=5.25cm}}
\centerline{(a)\hspace{5cm}(b)\hspace{5cm}(c)}
\caption{Simulated current histories for
non-uniform height of CNTs at a bias
voltage of (a) 440 V, (b) 550 V, and (c) 660 V.}
\label{fig_height}
\end{figure}

In this case, the uniform radius and the uniform spacing between neighboring CNTs are taken
as $r=2.75 nm$ and $d_1=2 \mu m$, respectively. Random initial distribution of the height is
given with bounds $10 - 14 \mu m$. The simulated results are shown in Fig.~\ref{fig_height}.
At the bias voltage of $440 V$, the average current decreases from $1.79 \times 10^{-6}A$ to
$1.53 \times 10^{-6}A$. The maximum current varies between $6.33 \times 10^{-6}A$ to $5.89
\times 10^{-6}A$, whereas the minimum current varies between $2.69 \times 10^{-10}A$ to $4.18
\times 10^{-10}A$. The average current decreases from $0.495 \times 10^{-3}A$ to $0.415
\times 10^{-3}A$ in steps when the bias voltage is increased from $440 V$ to $550 V$. At the
bias voltage of $660 V$, the average current decreases from $0.0231A$ to $0.0178A$. The
device response is found to be highly sensitive to the height distribution.

\subsubsection{Effects of non-uniform spacing between neighboring CNTs}
\begin{figure}
\centerline{\psfig{file=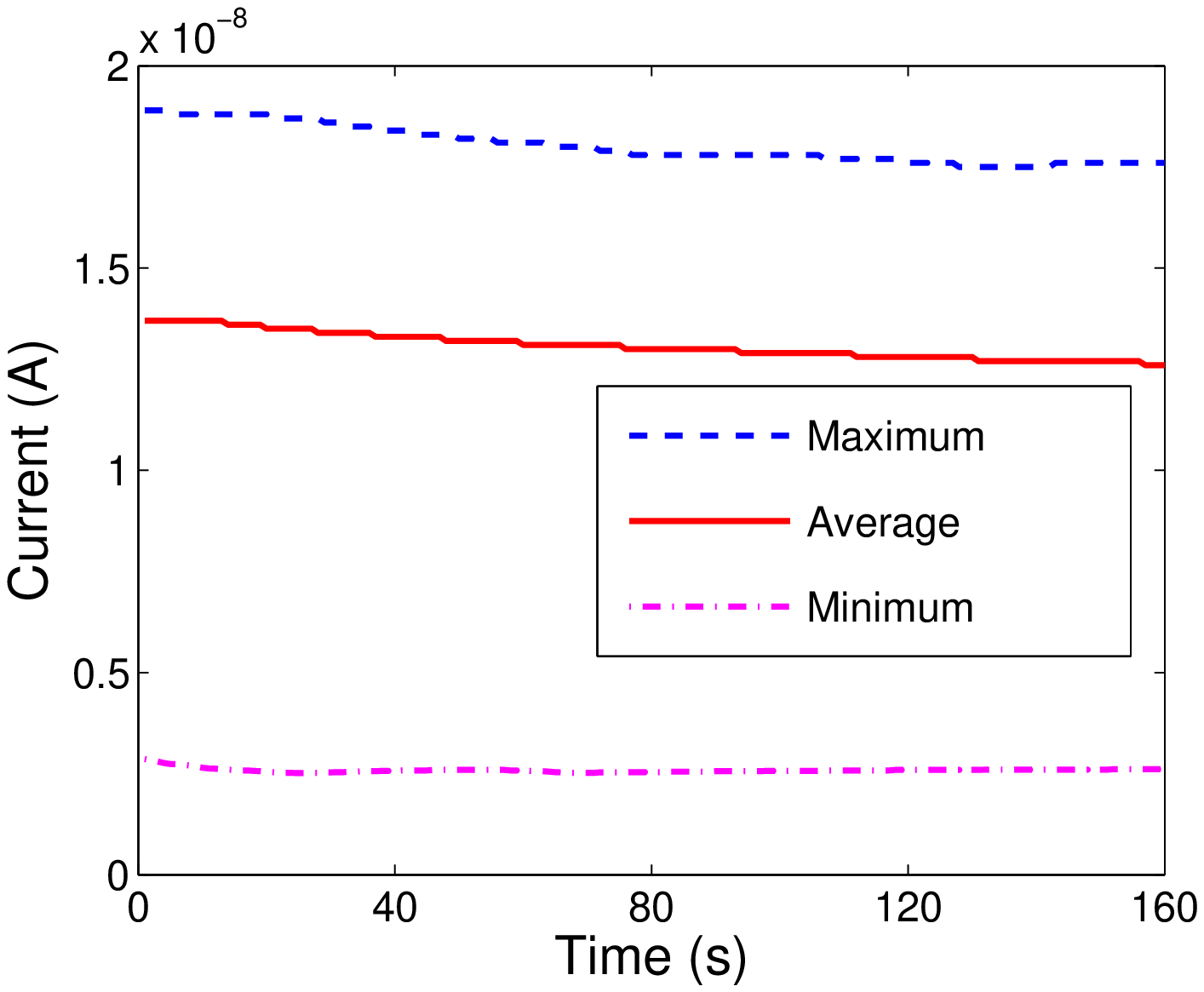,width=5.25cm}\hfill
            \psfig{file=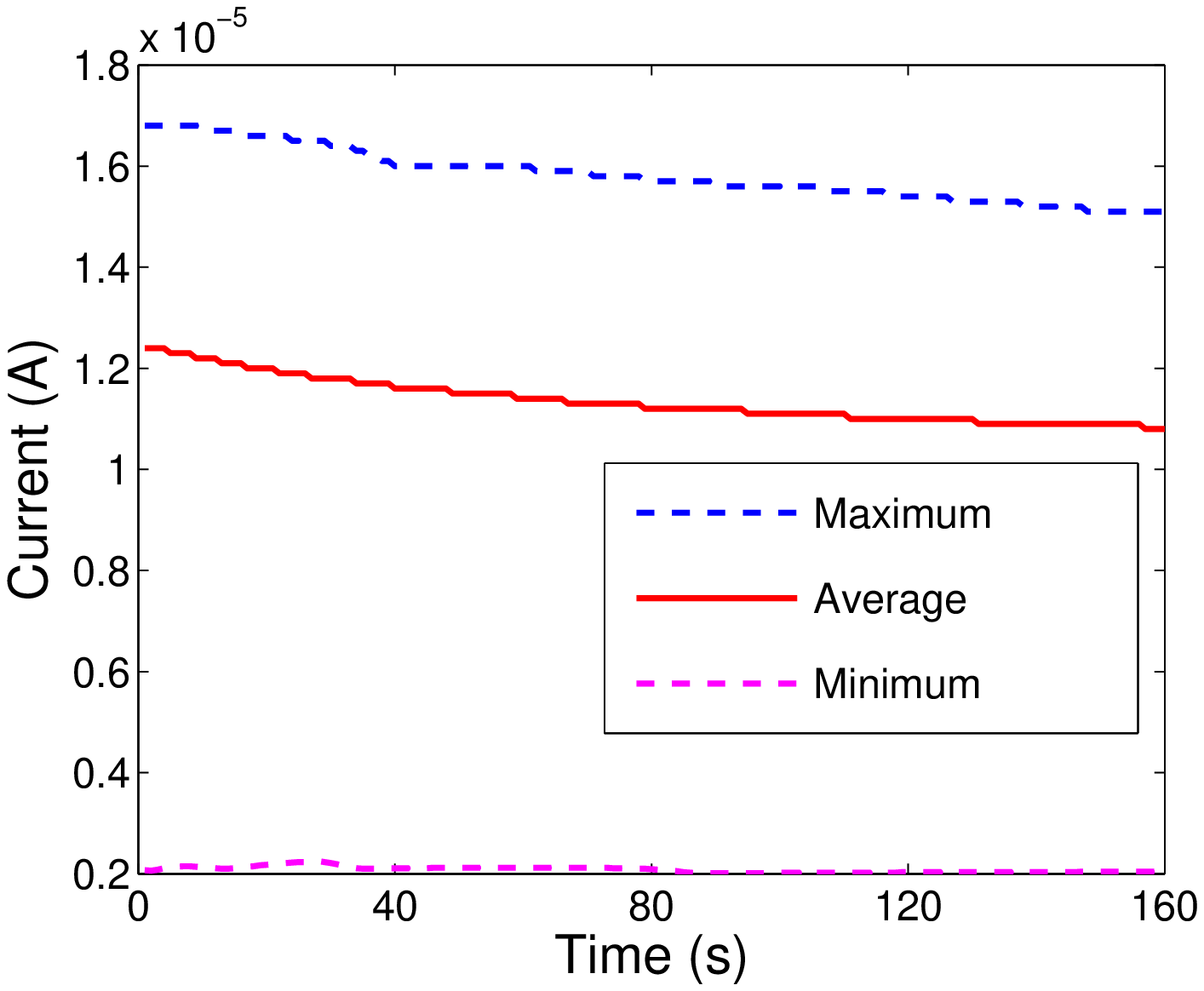,width=5.25cm}\hfill
            \psfig{file=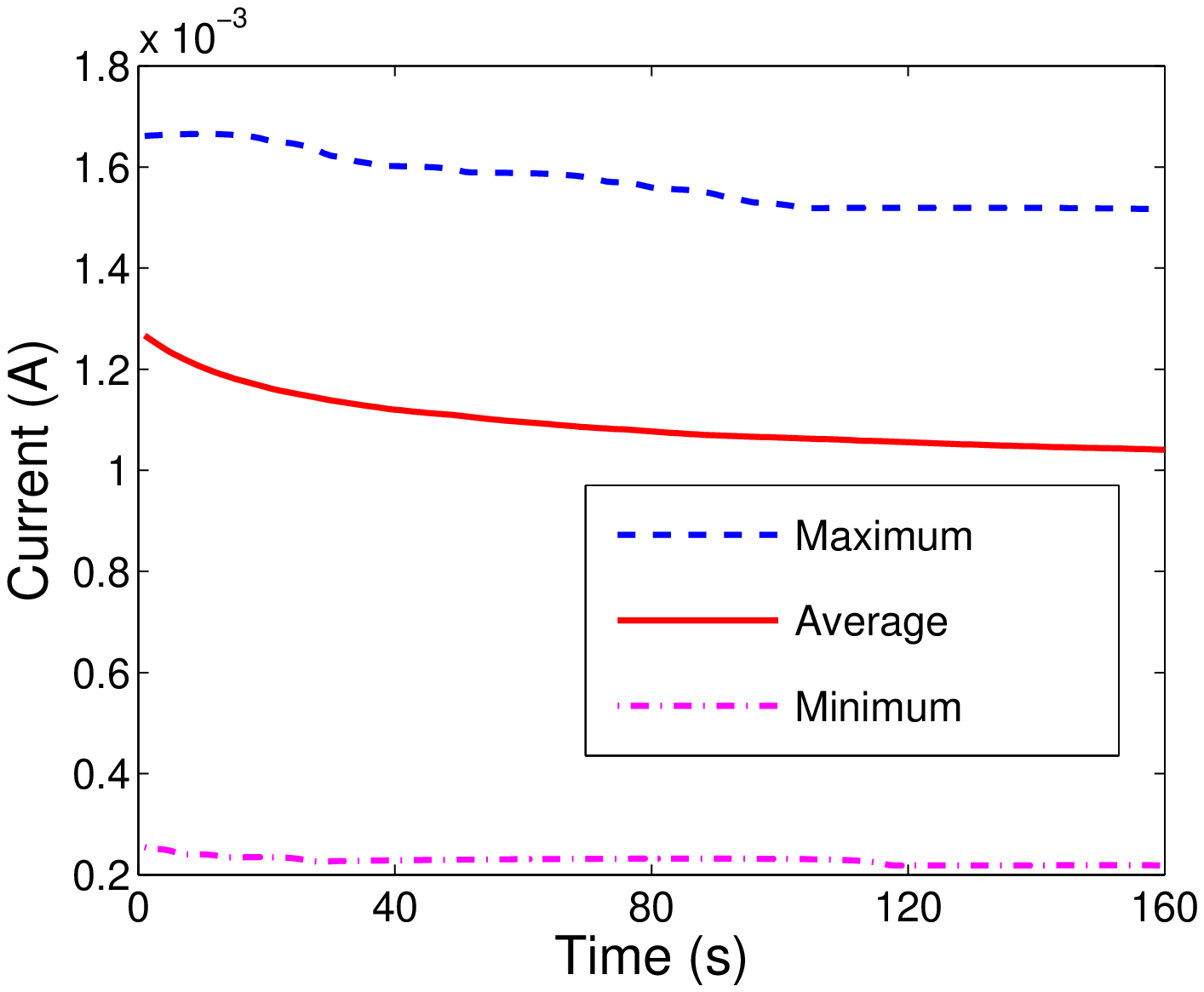,width=5.25cm}}
\centerline{(a)\hspace{5cm}(b)\hspace{5cm}(c)}
\caption{Simulated current histories for
non-uniform spacing between neighboring CNTs at a bias
voltage of (a) 440 V, (b) 550 V, and (c) 660 V.}
\label{fig_spacing}
\end{figure}

In this case, the uniform height and the uniform radius of the CNTs are taken as $h_0=12 \mu
m$ and $r=2.75 nm$, respectively. Random distribution of spacing $d_1$ between the
neighboring CNTs is given with bounds $1.5 - 2.5 \mu m$. The simulated results are shown in
Fig.~\ref{fig_spacing}. At the bias voltage of $440 V$, the average current decreases from
$1.37 \times 10^{-8}A$ to $1.26 \times 10^{-8}A$. The maximum current varies between $1.89
\times 10^{-8}A$ to $1.76 \times 10^{-8}A$, whereas the minimum current varies between $2.86
\times 10^{-9}A$ to $2.61 \times 10^{-9}A$. The average current decreases from $1.24 \times
10^{-5}A$ to $1.08 \times 10^{-5}A$ in steps when the bias voltage is increased from $440V$
to $550 V$. At the bias voltage of $660 V$, the average current decreases from $1.266 \times
10^{-3}A$ to $1.040 \times 10^{-3}A$. There is a slight increase in the order of magnitude of
current for non-uniform spacing. It can attributed to the reduction in screening effect at
some emitting sites in the film where the spacing is large.

\section{Conclusions}
\label{sec:concl} In this paper, we have developed a multiphysics based modelling approach to
analyze the evolution of the CNT thin film. The developed approach has been applied to the
simulation of the current-voltage characteristics at the device scale. First, a
phenomenological model of degradation and fragmentation of the CNTs has been derived. From
this model we obtain degraded state of CNTs in the film. This information, along with
electromechanical force, is then employed to update the initially prescribed distribution of
CNT geometries in a time incremental manner. Finally, the device current is computed at each
time step by using the semi-empirical Fowler-Nordheim equation and integration over the
computational cell surfaces on the anode side. The model thus handles several important
effects at the device scale, such as fragmentation of the CNTs, formation of the carbon
clusters, and self-assembly of the system of CNTs during field emission. The consequence of
these effects on the I-V characteristics is found to be important as clearly seen from the
simulated results which are in close agreement with experiments. Parametric studies reported
in the concluding part of this paper indicate that the effects of the height of the CNTs and
the spacing between the CNTs on the current history is significant at the fast time scale.

There are several other physical factors, such as the thermoelectric heating, interaction
between the cathode substrate and the CNTs, time-dependent electronic properties of the CNTs
and the clusters, ballistic transport etc., which may be important to consider while
improving upon the model developed in the present paper. Effects of some of these factors
have been discussed in the literature before in the context of isolated CNTs, but little is
known at the system level. We note also that in the present model, the evolution mechanism is
not fully coupled with the electromechanical forcing mechanism. The incorporation of the
above factors and the full systematic coupling into the modelling framework developed here
presents an appealing scope for future work.

\noindent
{\bf Acknowledgment}
The authors would like to thank Natural Sciences
and Engineering Research Council (NSERC), Canada, for financial support.

\newpage
\begin{table}
\begin{center}
\caption{Summary of work function values for CNTs.}
\begin{small}
\begin{tabular}{|l|l|l|}
\hline
Type of CNT & $\Phi$ ($eV$) & Method  \\
\hline SWNT & 4.8 & Ultraviolet photoelectron spectroscopy$^{45}$ 
\\
\hline MWNT & 4.3 & Ultraviolet photoelectron spectroscopy$^{46}$ 
\\
\hline MWNT & 7.3$\pm$0.5 & Field emission electronic
energy distribution$^{47}$ 
\\
\hline SWNT & 5.05 & Photoelectron emission$^{48}$ 
\\
\hline MWNT & 4.95 & Photoelectron emission$^{48}$ 
\\
\hline SWNT & 1.1 & Experiments$^{49}$ 
\\
\hline MWNT & 1.4 & Experiments$^{49}$ 
\\
\hline MWNT & 0.2-1.0 & Numerical approximation$^{50}$ 
\\
\hline
\end{tabular}
\label{table_workfunction}
\end{small}
\end{center}
\end{table}

\newpage

\newpage
\begin{thebibliography}{99}

\bibitem{bib:spindt}
C. A. Spindt, I. Brodie, L. Humphrey and E. R. Westerberg, 
J. Appl. Phys. 47, 5248 (1976).

\bibitem{bib:gotoh}
Y. Gotoh, M. Nagao, D. Nozaki, K. Utsumi, K. Inoue, T. Nakatani,
T, Sakashita, K. Betsui, H. Tsuji and J. Ishikawa, 
J. Appl. Phys. 95, 1537 (2004).

\bibitem{bib:zhu1}
W. Zhu (Ed.), Vacuum microelectronics, Wiley, NY (2001).

\bibitem{bib:iijima}
S. Iijima, 
Nature 354, 56 (1991).

\bibitem{bib:rinzler}
A. G. Rinzler, J. H. Hafner, P. Nikolaev, L. Lou, S. G. Kim, D. Tomanek,
D. Colbert and R. E. Smalley, 
Science 269, 1550 (1995).

\bibitem{bib:heer}
W. A. de Heer, A. Chatelain and D. Ugrate, 
Science 270, 1179 (1995).

\bibitem{bib:chernoz}
L. A. Chernozatonskii, Y. V. Gulyaev, Z. Y. Kosakovskaya, N. I. Sinitsyn,
G. V. Torgashov, Y. F. Zakharchenko, E. A. Fedorov and V. P. Valchuk,
Chem. Phys. Lett. 233, 63 (1995).

\bibitem{bib:bonard}
J. M. Bonard, J. P. Salvetat, T. Stockli, L. Forro and A. Chatelain,
Appl. Phys. A 69, 245 (1999).

\bibitem{bib:sugie}
H. Sugie, M. Tanemure, V. Filip, K. Iwata, K. Takahashi and F. Okuyama,
Appl. Phys. Lett. 78, 2578 (2001).

\bibitem{bib:groening}
O. Groening, O. M. Kuettel, C. Emmenegger, P. Groening, and L. Schlapbach,
J. Vac. Sci. Tech. B18, 665 (2000).

\bibitem{bib:fowler}
R. H. Fowler, and L. Nordheim, 
Proc. Royal Soc. London A 119, 173 (1928).

\bibitem{bib:bonard4}
J. M. Bonard, F. Maier, T. Stockli, A. Chatelain, W. A. de Heer, 
J. P. Salvetat and L. Forro, 
Ultramicroscopy 73, 7 (1998).

\bibitem{bib:dean1}
K. A. Dean, T. P. Burgin and B. R. Chalamala, 
Appl. Phys. Lett. 79, 1873 (2001).

\bibitem{bib:wei}
Y. Wei, C. Xie, K. A. Dean and B. F. Coll, 
Appl. Phys. Lett. 79, 4527 (2001).

\bibitem{bib:wang2}
Z. L. Wang, R. P. Gao, W. A. de Heer and P. Poncharal, 
Appl. Phys. Lett. 80, 856 (2002).

\bibitem{bib:chung}
J. Chung, K. H. Lee, J. Lee, D. Troya, and G. C. Schatz, 
Nanotechnology 15, 1596 (2004).

\bibitem{bib:avouris}
P. Avouris, R. Martel, H. Ikeda, M. Hersam, H. R. Shea and A. Rochefort, 
in Fundamental Mater. Res. Series, 
M. F. Thorpe (Ed.), Kluwer Academic/Plenum Publishers (2000) pp.223-237.

\bibitem{bib:nilsson1}
L. Nilsson, O. Groening, P. Groening and L. Schlapbach,
Appl. Phys. Lett. 79, 1036 (2001).

\bibitem{bib:nilsson2}
L. Nilsson, O. Groening, P. Groening and L. Schlapbach, 
J. Appl. Phys. 90, 768 (2001).

\bibitem{bib:bonard1}
J. M. Bonard, C. Klinke, K. A. Dean and B. F. Coll, 
Phys. Rev. B 67, 115406 (2003).

\bibitem{bib:bonard2}
J. M. Bonard, N. Weiss, H. Kind, T. Stockli, L. Forro, K. Kern 
and A. Chatelain, 
Adv. Matt. 13, 184 (2001).

\bibitem{bib:bonard3}
J. M. Bonard, J. P. Salvetat, T. Stockli, W. A. de Heer, L. Forro
and A. Chatelain, 
Appl. Phys. Lett. 73, 918 (1998).

\bibitem {bib:lim}
S. C. Lim, H. J. Jeong, Y. S. Park, D. S. Bae, Y. C. Choi, Y. M. Shin,
W. S. Kim, K. H. An and Y. H. Lee, 
J. Vac. Sci. Technol. A 19, 1786 (2001).

\bibitem{bib:huang2}
N. Y. Huang, J. C. She, J. Chen, S. Z. Deng, N. S. Xu, H. Bishop, S. E. Huq,
L. Wang, D. Y. Zhong, E. G. Wang and D. M. Chen, 
Phys. Rev. Lett. 93, 075501 (2004).

\bibitem{bib:zhu}
X. Y. Zhu, S. M. Lee, Y. H. Lee and T. Frauenheim, 
Phys. Rev. Lett. 85, 2757 (2000).

\bibitem{bib:collins}
P. G. Collins and A. Zettl, 
Appl. Phys. Lett. 69, 1969 (1996).

\bibitem{bib:nicol1}
D. Nicolaescu, L. D. Filip, S. Kanemaru and J. Itoh,
Jpn. J. Appl. Phys. 43, 485 (2004).

\bibitem{bib:nicol2}
D. Nicolaescu, V. Filip, S. Kanemaru and J. Itoh,
J. Vac. Sci. Technol. 21, 366 (2003).

\bibitem{bib:cheng}
Y. Cheng and O. Zhou, Electron field emission from carbon nanotubes,
C. R. Physique 4, 1021 (2003).

\bibitem{bib:sinha2}
N. Sinha, D. Roy Mahapatra, J. T. W. Yeow, R. V. N. Melnik 
and D. A. Jaffray, 
Proc. 6th IEEE Conf. Nanotech., Cincinnati, USA, July 16-20 (2006).

\bibitem{bib:sinha3}
N. Sinha, D. Roy Mahapatra, J. T. W. Yeow, R. V. N. Melnik and
D. A. Jaffray, 
Proc. 7th World Cong. Comp. Mech., Los Angeles, USA, 
July 16-22 (2006).

\bibitem{bib:fried1}
S. K. Friedlander, 
Ann. N.Y. Acad. Sci. 404, 354 (1983).

\bibitem{bib:girshick}
S. L. Grishick, C. P. Chiu and P. H. McMurry, 
Aerosol Sci. Technol. 13, 465 (1990).

\bibitem{bib:jiang}
H. Jiang, P. Zhang, B. Liu, Y. Huang, P. H. Geubelle, H. Gao and
K. C. Hwang, 
Comp. Mat. Sci. 28, 429 (2003).

\bibitem{bib:slepyan}
G. Y. Slepyan, S. A. Maksimenko, A. Lakhtakia, O. Yevtushenko
and A. V. Gusakov, 
Phys. Rev. B 60, 17136 (1999).

\bibitem{bib:xiao}
J. R. Xiao B. A. Gama and J. W. Gillespie Jr.,
Int. J. Solids Struct. 42, 3075 (2005).

\bibitem{bib:ruoff}
R. S. Ruoff, J. Tersoff, D. C. Lorents, S. Subramoney and B. Chan,
Nature 364, 514 (1993).

\bibitem{bib:hertel}
T. Hertel, R. E. Walkup and P. Avouris, 
Phys. Rev. B 58, 13870 (1998).

\bibitem{bib:glukhova}
O. E. Glukhova, A. I. Zhbanov, I. G. Torgashov, N. I. Sinistyn and
G. V. Torgashov, Appl. Surf. Sci. 215, 149 (2003).

\bibitem{bib:musatov}
A. L. Musatov, N. A. Kiselev, D. N. Zakharov, E. F. Kukovitskii,
A. I. Zhbanov, K. R. Izrael'yants and E. G. Chirkova,  
Appl. Surf. Sci. 183, 111 (2001).

\bibitem{bib:huang}
Z. P. Huang, Y. Tu, D. L. Carnahan and Z. F. Ren,
in Encycl. Nanosci. Nanotechnol. 3, Edited by H. S. Nalwa, 
American Scientific Publishers, Los Angeles (2004), pp.401-416.

\bibitem{bib:gadzuk}
J. W. Gadzuk and E. W. Plummer,  
Rev. Mod. Phys. 45, 487 (1973).

\bibitem{bib:dean3}
K. A. Dean, O. Groening, O. M. Kuttel and L. Schlapbach, 
Appl. Phys. Lett. 75, 2773 (1999).

\bibitem{bib:takakura}
A. Takakura, K. Hata, Y. Saito, K. Matsuda, T. Kona and C. Oshima, 
Ultramicroscopy 95, 139 (2003).

\bibitem{bib:suzuki}
S. Suzuki, C. Bower, Y. Watanabe and O. Zhou, 
Appl. Phys. Lett. 76, 4007 (2000).

\bibitem{bib:ago}
H. Ago, T. Kugler, F. Cacialli, W. R. Salaneck, M. S. P. Shaffer, A. H. Windle
and R. H. Friend, 
J. Phys. Chem. B 103, 8116 (1999).

\bibitem{bib:fransen}
M. J. Fransen, T. L. van Rooy and P. Kruit, 
Appl. Surf. Sci. 146, 312 (1999).

\bibitem{bib:shiraishi}
M. Shiraishi and M. Ata, 
Carbon 39, 1913 (2001).

\bibitem{bib:sinitsyn}
N. I. Sinitsyn, Y. V. Gulyaev, G. V. Torgashov, L. A. Chernozatonskii, 
Z. Y. Kosakovskaya, Y. F. Zakharchenko, N. A. Kiselev, A. L. Musatov, 
A. I. Zhbanov, S. T. Mevlyut and O. E. Glukhova, 
Appl. Surf. Sci. 111, 145 (1997).

\bibitem{bib:obraztsov}
A. N. Obraztsov, A. P. Volkov and I. Pavlovsky, 
Diam. Rel. Mater. 9, 1190 (2000).

\end{thebibliography}
\end{document}